\documentclass{ar2e}

\usepackage{natbib}
\bibpunct{(}{)}{,}{a}{}{,}

\begin{document}
\def\s{{\rm\,s}} 
\def\sr{{\rm\,sr}} 
\def\erg{{\rm\,erg}} 
\def\cm{{\rm\,cm}} 
\def\m{{\rm\,m}} 
\def\km{{\rm\,km}} 
\def\mm{{\rm\,mm}} 
\def\gm{{\rm\,g}} 
\def\g{{\rm\,g}} 
\def\kg{{\rm\,kg}} 
\def\au{{\rm AU}} 
\def\deg{{\rm deg}} 
\def\rad{{\rm rad}}   
\def\AU{{\rm\, AU}}  
\def\Ri{{\rm Ri}}  
\def\Reo{{\rm Re}_{\rm o}} 
\def\Rey{{\rm Re}}  
\def\K{{\rm\,K}}  
\def\yr{{\rm\,yr}}  
\def\Hz{{\rm\,Hz}}  
\def\mum{\mu{\rm m}}  
\def\ts{t_{\rm s}}
\def\ti{t_{\rm i}}
\def\si{s_{\rm i}}
\def\to{t_{\rm o}}
\def\vo{\delta v_{\rm o}}
\def\vp{\delta v_{\rm p}}
\def\vpp{\delta v_{\rm p,p}}

\def\hp{h_{\rm p}}
\def\hg{h_{\rm g}}
\def\lo{l_{\rm o}}
\def\ello{\ell_{\rm o}}
\def\vell{v_\ell}
\def\li{l_{\rm i}}
\def\elli{\ell_{\rm i}}
\def\vi{\delta v_{\rm i}}
\def\Dp{D_{\rm p}}
\def\Dg{D_{\rm g}}
\def\OmegaK{\Omega_{\rm K}}
\def\vk{v_{\rm K}}
\def\taus{\tau_{\rm s}}
\def\tauo{\tau_{\rm o}}
\def\St{{\rm St}}
\def\rhos{\rho_{\rm s}}
\def\rhog{\rho_{\rm g}}
\def\rhop{\rho_{\rm p}}
\def\rhopg{\langle\rho_{\rm p}/\rho_{\rm g}\rangle}
\def\Sigmap{\Sigma_{\rm p}}
\def\Sigmag{\Sigma_{\rm g}}
\def\cg{c_{\rm g}}
\def\eV{{\rm\,eV}}  
\def\baru{{\rm\,bar}}
\def\lesssim{\mathrel{\hbox{\rlap{\hbox{\lower4pt\hbox{$\sim$}}}\hbox{$<$}}}}
\def\gtrsim{\mathrel{\hbox{\rlap{\hbox{\lower4pt\hbox{$\sim$}}}\hbox{$>$}}}}
\def\ion#1#2{#1$\;${\small\rm\@{#2}}\relax}
\def\Zr{Z_{\rm rel}}
\def\sigmol{\sigma(\rm H_2)}
\newcommand{\p}{\partial}
\newcommand{\vc}[1]{\mbox{\boldmath{$#1$}}}
\newcommand{\Eq}[1]{equation\,(\ref{#1})}

\input epsf.def   

\input psfig.sty

\jname{Annual Reviews of Earth and Planetary Science}
\jyear{2009}
\jvol{38}
\ARinfo{1056-8700/97/0610-00}

\title{FORMING PLANETESIMALS IN SOLAR AND EXTRASOLAR NEBULAE}

\renewcommand{\baselinestretch}{1.00}

\markboth{CHIANG \& YOUDIN---SUBMITTED DRAFT}{PLANETESIMAL FORMATION---SUBMITTED DRAFT}

\author{E. Chiang$^{1,2}$ and A. Youdin$^3$ \affiliation{1. Department
    of Astronomy, 601 Campbell Hall, University of California,
    Berkeley CA 94720, USA; email: echiang@astro.berkeley.edu \\
    2. Department of Earth and Planetary Science, 307 McCone Hall,
    University of California, Berkeley CA 94720, USA \\ 3. Canadian
    Institute for Theoretical Astrophysics, University of Toronto, 60
    Saint George St., Toronto ON M5S 3H8, Canada; email:
    youd@cita.utoronto.ca}}

\begin{keywords}
planet formation, Solar System, accretion, planets, circumstellar disks, fluid mechanics, turbulence
\end{keywords}

\begin{abstract}

  Planets are built from planetesimals: solids larger than a kilometer
  which grow by colliding pairwise.  Planetesimals themselves are
  unlikely to form by two-body collisions; sub-km objects have
  gravitational fields individually too weak, and electrostatic
  attraction is too feeble for growth beyond a few cm.  We review the
  possibility that planetesimals form when self-gravity brings
  together vast ensembles of small particles.  Even when self-gravity
  is weak, aerodynamic processes can accumulate solids relative to
  gas, paving the way for gravitational collapse.  Particles pile up
  as they drift radially inward.  Gas turbulence stirs particles, but
  can also seed collapse by clumping them. While the feedback of
  solids on gas triggers vertical shear instabilities that obstruct
  self-gravity, this same feedback triggers streaming instabilities
  that strongly concentrate particles.  Numerical simulations find
  that solids $\sim$10--100 cm in size gravitationally collapse in
  turbulent disks. We outline areas for progress, including the
  possibility that still smaller objects self-gravitate.
\end{abstract}

\maketitle

\small\normalsize
\parskip = 0.5em

\section{INTRODUCTION}\label{sec_intro}
\begin{figure}
\centerline{\psfig{figure=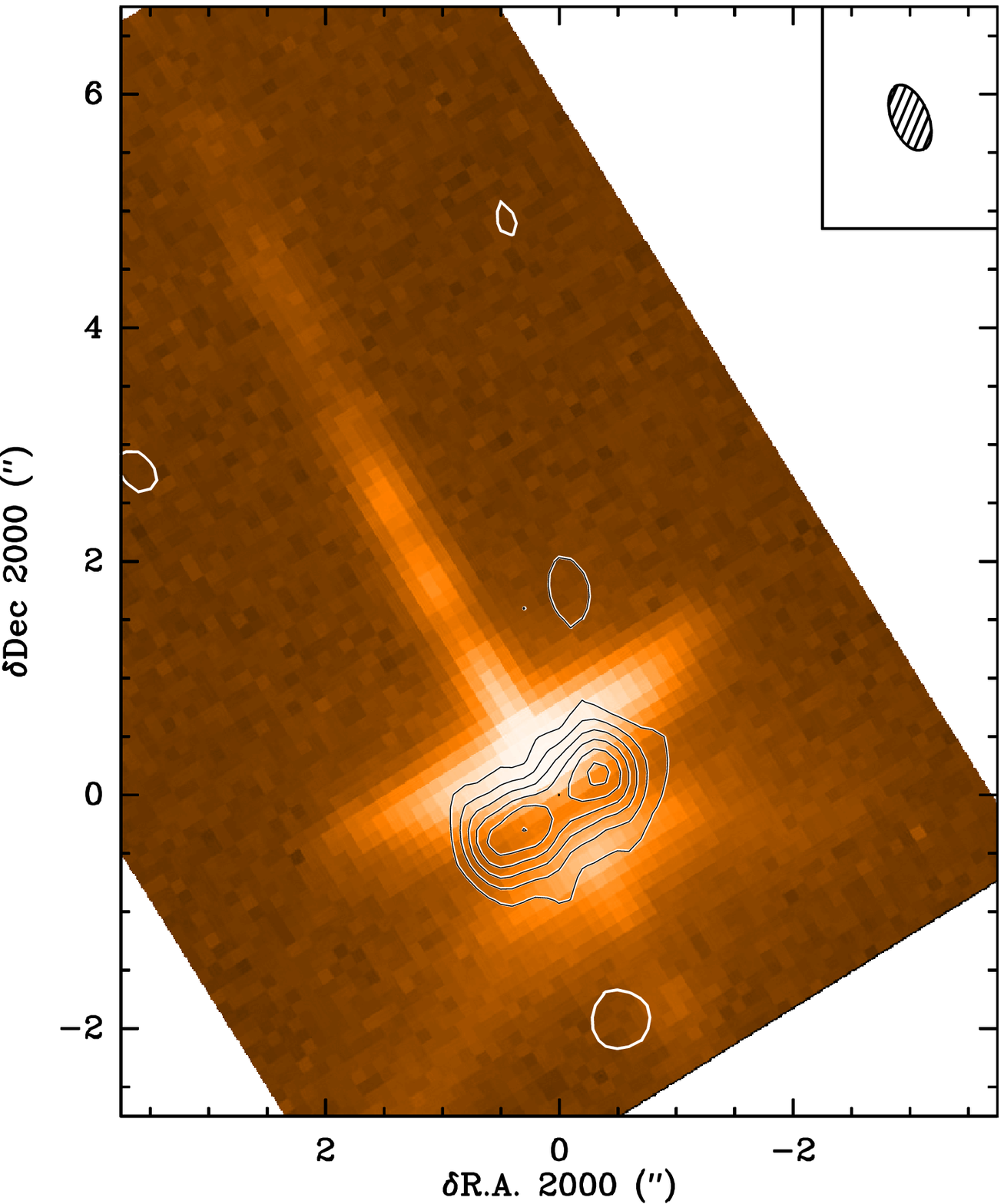,height=25pc}}
\caption{Hubble Space Telescope image of the protoplanetary disk HH
  30, seen in scattered starlight at optical wavelengths
  \citep{burrows}.  Overlaid are contours from 1.3-mm continuum
  radiation measured by the Plateau de Bure Inteferometer
  \citep[PdBI,][]{guilloteau}. The optical image reveals the opaque
  disk midplane, a dark lane dividing two ``bowls'' of light scattered
  by sub-$\mum$ grains in flared disk surfaces.  Perpendicular to the
  disk plane is the jet.  Mm-wave emission arises predominantly from
  the midplane, plausibly from cm-sized grains that have settled
  vertically. The mm-wave disk has an outer radius of $\sim$130 AU,
  smaller than the optical radius of $\sim$300 AU, suggesting either
  that grains grow faster at smaller radius or that cm-sized grains
  have drifted radially inward (\S\ref{sec_lam}). Axes are measured in
  arcseconds; 1$''$ = 140 AU for an assumed distance of 140 parsecs.
  The inset shaded spot shows the size of the PdBI beam, a measure of
  the angular resolution of the mm-wave map. Reproduced by permission
  of Astronomy \& Astrophysics.}
\label{fig:HH30}
\end{figure}

The nebular hypothesis---that planets coalesce from disks of gas and
dust orbiting young stars---is confirmed today in broad
outline. Protoplanetary disks of ages 1--10 Myr are now studied
routinely (Figure \ref{fig:HH30}; \citealt{watson}). They contain enough
mass to spawn planets like those detected around hundreds of Gyr-old
stars \citep{jones}. Debris disks, observable via dust
generated from collisions between larger parent bodies, bridge our understanding
in the 10--100 Myr interval
\citep{wyatt}.  These parent bodies may represent
first-generation planetesimals, the building blocks of planets. Our
Solar System may preserve a record of planet formation, in the size
distributions of asteroids \citep{morbi09} and Kuiper belt
objects \citep{pan}.

The remarkable journey that disk solids make in growing from
microscopic dust to Earth-mass ($M_\oplus$) planets divides into three
legs. At the smallest sizes $\lesssim \cm$, chemical bonds and van der
Waals forces enable grains to stick to one another. At the largest
sizes $\gg \km$, gravity promotes growth. Pairs of objects coagulate
upon colliding because their gravity is strong enough to retain
collision fragments.

At intermediate sizes lies the domain of planetesimal formation.
It presents the most challenging of terrains; here grain surfaces
are insufficiently sticky and the gravity of a single object too
feeble. \citet{saf} kindled the hope that where the individual fails,
the collective might succeed: if solid particles have a large enough
density en masse, their gravity can draw them together. The many
calculations seeking to realize this vision are the subject of our
review.

An incompressible fluid body in hydrostatic equilibrium, orbiting a
star of mass $M_{\ast}$ at distance $r$, has sufficient self-gravity to
resist disruption by the star's tidal gravitational field if its
density exceeds the Roche value $\rho_{\rm Roche} \approx 3.5
M_{\ast}/r^3$ \citep{chandrasek87}. Though we will see that the
effects of self-gravity manifest at arbitrarily low densities
(\S\ref{sec:draggi}), the Roche density is a fine benchmark against
which to measure progress towards forming planetesimals, and we will
use it as such. It is a formidably large density, 2--3 orders of
magnitude greater than typical densities at disk midplanes
(\S\ref{sec_disks}). \citet{saf}, and independently \citet[][hereafter
GW]{gw}, proposed that densities approaching Roche (actually a factor
of 20 smaller, see \S\ref{sec:dyncollapse}) could be achieved by having dust
settle vertically and accumulate in a thin ``sublayer'' at the
midplane. \citet{weiden80} pointed out that large densities were not
so easily achieved---that turbulence generated by vertical shear
across the sublayer would halt settling and prevent gravitational
instability (GI).  GW appreciated that the sublayer would be
turbulent, but overlooked the decisive impact of turbulence on the
layer's thickness and density.

This stalemate has been broken---indeed the entire playing field
redrawn---in the last decade by abandoning three assumptions. First,
the height-integrated surface density ratio of dust to gas need not be
solar. Enriching the disk in metals, say by photoevaporation of gas
\citep{tb05, gorti, ercolano} or by radial drifts of particles \citep[][hereafter
YS]{ys}, can stop vertical shear turbulence from forestalling GI
(\citealt{sek98}, YS, \citealt{c08}). Second, turbulence, of whatever
origin, is not always the enemy of GI. Magneto-rotational tubulence is
observed in numerical simulations to generate long-lived structures in
gas that can trap particles \citep{fn05,jkh06}. Third, gas drags
particles, but by Newton's Third Law, particles also drag
gas. Properly accounting for the backreaction of particles on gas
leads to powerful drag instabilities that can concentrate particles
and seed GI \citep{gp00,yg05,jy07}.

This review is organized as follows. Section 2 establishes
order-of-magnitude properties of protoplanetary disks using an
up-to-date model for the minimum-mass solar nebula. Attention is paid
to primitive chondritic meteorites. Section 3 gives a primer on the
aerodynamics of grains, treating gas drag in the ``test particle''
limit where backreaction is neglected. We discuss how grain
dynamics are affected by gas turbulence. Section 4 describes the
extent to which grains can grow by sticking. Section 5 rehearses and
comments on Toomre's criterion for GI---and how the criterion is
removed when self-gravity is combined with gas drag. Section 6 treats
sublayer shearing instabilities.  Section 7 introduces secular drag
instabilities that concentrate particles without recourse to
self-gravity. We offer new insights into Goodman \& Pindor's
\citeyearpar{gp00} toy model for drag instabilities. Section 8
outlines the nonlinear outcome of GI, describing breakthrough
numerical simulations. Finally, a summary is
supplied in Section 9, where we also list a few forefront
problems. Our study is peppered throughout with calculations
that readers are encouraged to reproduce---and improve.

Our review complements others in the field of planet
formation. \citet{blumwurm08} 
do more justice than we have to the
growth of particle aggregates by grain-grain sticking. \citet{g04}
provide a pedagogical and cutting-edge review of protoplanet accretion
by gravitationally focussed, pairwise collisions. For a
review of planetesimal formation that overlaps ours and
covers some topics in greater detail,
see the Les Houches lectures by \citet{houches}.

\section{PROTOPLANETARY DISKS}\label{sec_disks}

A starting point for calculations is the minimum-mass solar nebula
(MMSN), derived by adding enough H and He to solar system planets to
restore them to solar composition, and spreading the augmented masses
into abutting annuli centered on their orbits
\citep[e.g.,][]{weidenschilling77a}.  This exercise is uncertain, since
estimates of the total ``metal'' (non-H and non-He) content of Jupiter
range from 10 to 42 $M_\oplus$; for Saturn the range is 15 to 30
$M_\oplus$ \citep{guillot}. Even if these uncertainties were reduced,
it is unlikely our solar system formed strictly from the MMSN.  The conversion
of disk metals into planets cannot be 100\% efficient, and current
planetary orbits may differ significantly from their original ones
\citep[e.g.,][]{malhotra}.  Moreover a major theme of this review is
that planetesimal formation may require the metal fraction to evolve
to values greater than that of the bulk Sun, at least in certain
regions of the disk.  Thus the utility of the MMSN lies in providing a
baseline for discussing real-world complications, and in establishing
orders of magnitude.

For numerical estimates in this review, we adopt a disk surface
density:
\begin{eqnarray}
\label{eq_sigmag}
\Sigma_{\rm g} = 2200 \,F \left( \frac{r}{\rm AU} \right)^{-3/2} \gm \cm^{-2} \\
\label{eq_sigmad}
\Sigma_{\rm p} = 33 \,F\, \Zr \left( \frac{r}{\rm AU} \right)^{-3/2} \gm \cm^{-2}
\end{eqnarray}
where subscripts ${\rm g}$ and ${\rm p}$ denote gas and particles
(condensed metals), respectively.  Our MMSN  ($F=1$,
$\Zr=1$),  uses the updated condensate mass fraction for Solar abundances of 0.015
\citep{lodders}.  When methane ice sublimates above temperature $T
\approx 41 \K$, $\Zr = 0.78$; above $\sim$$182\K$, water and all other
ices are lost so that $\Zr = 0.33$ \citep{lodders}.  The coefficient for
eq.~(\ref{eq_sigmad}) is chosen to give $1 M_\oplus$ of solids in an
annulus centered on the Earth's orbit when $F = 1$ and $\Zr = 0.33$.  We
invoke values of $\Zr > 1$ to account for various metal enrichment
processes, depicted in Figure \ref{fig:enrich} and discussed throughout this review
(e.g., \citealt{sek98}, YS, \citealt{tb05}).

Integrated to $r = 100\AU$, eq.~(\ref{eq_sigmag}) yields a mass of
$0.03 F M_{\odot}$.  Astronomical observations of disks orbiting
Myr-old, Sun-like stars suggest they contain 0.001--0.1 $M_{\odot} =
1$--100 $M_{\rm J}$ of gas and dust \citep[][]{aw05}.  Disk masses are
derived from mm-wave radiation from dust, located $\sim$100 AU from
host stars. From the dust emission is calculated a dust mass, and from
the dust mass a gas mass is extrapolated assuming a solar dust-to-gas
ratio. The gas mass thus imputed is uncertain because the disk
metallicity is not known, and because the modeled dust mass depends on
an unknown grain size distribution. Disk masses may be systematically
underestimated because grains may have grown to sizes $\gg$ mm and
would therefore be practically invisible at sub-cm wavelengths
\citep[e.g.,][]{hartmann}.

Based on their near-ultraviolet excess emission \citep{calvet98},
young Sun-like stars accrete gas from disks at a typical rate
$\dot{M}_{\ast} \sim 10^{-8} M_{\odot} \yr^{-1}$
\citep[e.g.,][]{hartmann}. Accretion implies that disk gas cannot
everywhere be static (``passive''). Mass is transported inward by the
outward transport of angular momentum, either by turbulence or by
ordered flows \citep{fkr}. The transport mechanism---or, as is
commonly stated, the origin of disk viscosity---remains obscure, with
proposals ranging from turbulence driven by the magneto-rotational
instability (MRI; \citealt{balbus09}) to vortices
\citep{lithwick} to gravitational torques \citep{vb4}.
When transport is local, our ignorance is encapsulated in the
parameter $\alpha$ \citep[e.g.,][]{fkr}, the ratio of the local shear
stress to the total pressure.  Crude, disk-averaged values of $\alpha
\sim 10^{-2}$ are inferred from observations of $\dot{M}_{\ast}$
vs. age \citep{calvet00}.

Accretion may be restricted to certain radii and times, as in models
that rely upon the MRI---not all of the disk may be sufficiently
ionized to couple to magnetic fields \citep[e.g.,][see also \citealt{cmc}]{bai}.
Magnetically inert regions are called ``dead zones'' \citep{gammie96}.
Whether planetesimals form in active or passive regions of the disk is
an outstanding question. This review will explore both cases.

The fraction of stars with near-infrared excess emission attributable
to optically thick, gaseous disks decreases from near unity at stellar
ages $\lesssim 1$ Myr, to $\lesssim 5$\% at $\gtrsim 10$ Myr
\citep{hern,hill05}. 
Whatever combination of accretion, photoevaporation, and planet
formation is responsible for this observed decline, giant
planets---including ice giants, which contain more hydrogen than can
be explained by accretion of hydrated solids \citep{liss07}---must
form within several Myr before the gas dissipates.  The same deadline
characterizes planetesimal formation, since the ice giants, which by
mass are 80--90\% rock and ice, coagulated from planetesimals.

Current models set the formation of solar system giant planets within
a disk with more solids, and sometimes more gas, than the
MMSN. \citet{g04} identify Neptune and Uranus with ``isolation-mass''
oligarchs at $r \approx 20$--30 AU, requiring $F \Zr\approx
6$.\footnote{A protoplanet of mass $M_p$ becomes ``isolated'' when it
  has accreted an annulus of disk material having a width about
  $5\times$ the radius of its Hill sphere, $R_{\rm H} \approx
  [M_p/(3M_{\ast})]^{1/3}r$ \citep[e.g.,][]{greenberg}.  The
  average density of the planet spread through its Hill sphere is of
  order $\rho_{\rm Roche} \approx 3.5 M_{\ast}/r^3$.}
  \citet{liss09} prefer $F \Zr \approx 3.5$ at
$r\approx 5\AU$, so that Jupiter's core can accrete its gaseous
envelope within 3 Myr.

At the midplane of a passive disk---one heated solely by stellar
radiation---the gas temperature, scale height, and density are
approximately
\begin{eqnarray}
\label{eq_T}
T = 120 \left( \frac{r}{\rm AU} \right)^{-3/7} \K \\
\label{eq_h}
h_{\rm g} = 0.022 
r \left( \frac{r}{\rm AU} \right)^{2/7} \\
\label{eq_rho}
\rhog = 2.7  
\times 10^{-9} F \left( \frac{r}{\rm AU} \right)^{-39/14} \gm \cm^{-3} \,.
\end{eqnarray}
These are adapted from \citet{cg97}, adjusted for a disk obeying
(\ref{eq_sigmag})--(\ref{eq_sigmad}), orbiting a young
star of mass $M_{\ast} = 1 M_{\odot}$, radius $R_{\ast} = 1.7
R_{\odot}$, and temperature $T_{\ast} = 4350 \K$.
Thus in our MMSN, water ice starts to condense outside $\sim$$0.4$ AU,
and methane freezes outside $\sim$12 AU.
More detailed models of passive disks are reviewed by
\citet{dulle}. Turbulent accretion can give higher midplane
temperatures.  The density and thermal structure of active disks
depends on the assumed viscosity profile, $\alpha(r)$.
One model is that of \citet{daless}, which assumes $\alpha(r) = {\rm
  constant}$ and is tailored to fit broadband spectra and reflected
light images. For simplicity, estimates in this review employ
(\ref{eq_sigmag})--(\ref{eq_rho}) for both active and passive disks.

What do astronomical observations tell us about grain sizes? In disk
surface layers directly illuminated by optical light from host stars,
grain sizes are 1--$10 \, \mum$, as deduced from 
mid-infrared silicate emission bands \citep{natta}, and from scattered
light images at similar wavelength \citep{mccabe}.  Surface grains
have settled vertically, residing at heights $z$ above the midplane of
1--$3 h_{\rm g}$ \citep{c01}; for a disk in which dust and gas are well mixed,
$z/h_{\rm g} \approx 4$--5.  At disk midplanes, mm to cm-sized grains are
routinely invoked to match mm to cm-wave spectra and images
\citep[e.g.,][]{daless,testi}.

The most primitive meteorites, of nearly solar photospheric
composition except in volatile elements, also offer data on
particle sizes.
Confounding meteoriticists and astronomers alike is why up to
$\sim$90\% of their volume is filled with chondrules: once molten,
0.1-mm to cm-sized spheres that solidified 4.57 Gyr ago
\citep[e.g.,][]{hewins}.  Chondrule petrology is consistent with their
having been heated just above liquidus for less than minutes, and
having cooled for hours to days.  The heating mechanism is not known;
nebular shocks are suspected, but the origin of such shocks is debated
\citep{deschetal}.  Chondrules might well have been the building
blocks of the first-generation planetesimals, brought together by
self-gravity.  That $\sim$10\% of chondrules are binaries which
collided and fused while still partially molten implies that when
chondrules were suspended in space, they had large collective
densities \citep{goodingkeil}, possibly exceeding the Roche value.
Extremely dusty environments are also indicated by the retention of
volatiles in chondrules \citep{alex08}.

\section{AERODYNAMICS OF INDIVIDUAL PARTICLES}\label{sec_aero}
Gas drags particles. The degree of coupling is measured by the
dimensionless stopping time $\tau_{\rm s} \equiv \Omega_{\rm K} t_{\rm s}$, where
$\Omega_{\rm K}$ is the Keplerian angular velocity and
\begin{eqnarray} 
\ts \equiv m v_{\rm rel} / F_{\rm D} \approx \left\{ \begin{array}{ll}
        \rhos s / (\rhog \cg) & \mbox{if $s \lesssim 9 \lambda / 4$ (Epstein)} 
\\
     4  \rhos s^2 / (9 \rhog \cg \lambda) & \mbox{if $s \gtrsim 9 \lambda/4$, $ \Rey \lesssim 1$ (Stokes)} \label{eq_taus}
\end{array}
\right.
\end{eqnarray}
measures how long it takes a particle of mass $m$, radius $s$, and
internal density $\rhos$ to have its speed $v_{\rm rel}$ relative to
gas be reduced by order unity. The gas sound speed is $\cg$.
Particles are well entrained in gas when $\taus \ll 1$.  We have given
the two cases for the drag force $F_{\rm D}$ most relevant for
planetesimal formation: Epstein's \citeyearpar{epstein} law of free
molecular drag, arising from the difference $\sim$$\rhog [(\cg +
v_{\rm rel})^2 - (\cg - v_{\rm rel})^2]$ in momentum fluxes received
by the windward and leeward faces of the particle, and Stokes drag for
low Reynolds number $\Rey \equiv sv_{\rm rel}/(\lambda \cg)$. Given
(\ref{eq_rho}), the mean free path for collisions between gas
molecules is $\lambda \approx 0.5 F^{-1} (r/{\rm AU})^{39/14}
\cm$.\footnote{We use a constant molecular cross section, $\sigmol
  \approx 2 \times 10^{-15}\cm^2$, that reproduces the dynamical
  viscosity $\mu$ at 200 K.  But constant $\sigmol$ assumes $\mu
  \propto \sqrt{T}$, which fails for $T \lesssim 70$ K, the Sutherland
  constant for H$_2$.  At colder temperatures $\mu \propto T^{3/2}$ so
  that $\sigmol \propto T^{-1}$ \citep{chapcow}.  For simplicity we
  neglect the lower values of viscosity and $\lambda$ that this effect
  would produce in cold regions. \label{foot_chapcow}} Other cases for
$F_{\rm D}$ are given by \citet{ahn76} and \citet{weidenschilling77b}.
For $F = 2$ and $\rhos = 1~{\rm g/cm^3}$, marginally coupled bodies
($\taus = 1$) have $s = s_1$ increasing from 35 to 120 
cm in the Stokes regime as $r$ runs from 1 to 7 
AU; thereafter $s_1$ declines with $r$ in the Epstein regime, with
$s_1 \approx 13$ 
cm at 30 AU.

Gas and dust move at different velocities because pressure gradients
barely accelerate particles with $\rhos \gg \rhog$. For the moment,
let us neglect turbulence, and assume the collective particle density
$\rhop \ll \rhog$ so we can ignore the backreaction of dust on
gas. Then the azimuthal gas velocity $v_{{\rm g}\phi}$ obeys
\begin{equation}
\frac{v_{{\rm g}\phi}^2}{r} = \frac{GM_{\ast}}{r^2} - \frac{1}{\rhog} \frac{\partial P}{\partial r} \,.
\end{equation}
Pressure $P \approx \rhog \cg^2$ makes the gas rotate more slowly than the local Keplerian
velocity  $\vk \equiv \OmegaK r = \sqrt{GM_{\ast}/r}$ by
\begin{equation}\label{eq_etavK}
\eta \vk \equiv \vk - v_{{\rm g}\phi}  \approx - \frac{\p P /\p \ln r}{2 \rhog v_{\rm K}} 
\approx 25 \left(\frac{r}{\au}\right)^{1/14} \m \s^{-1}
\end{equation}
using the thin disk approximation $\cg/\vk = \hg/r \ll 1$.  While
$\eta \vk \sim \cg^2/\vk$ is independent of total disk mass and is
nearly independent of $r$, it increases in hotter disks, perhaps to
$\gtrsim 50 \m \s^{-1}$. An individual particle seeks to orbit at the
full $\vk$.  Thus it normally experiences a headwind, except possibly
where there are variations to the power-law descent of $P(r)$, as
considered at the end of \S\ref{sec_pile}.

\subsection{Particle Drifts and Consequences}\label{sec_drift}
\subsubsection{Laminar Drift Speeds}\label{sec_lam}
In a passive disk, a particle settles vertically to the midplane and,
on longer timescales, drifts radially inward as the headwind saps its
angular momentum. The following is drawn from \citet[][see also
\citealt{nakagawa}]{houches}. If we neglect backreaction, the
cylindrical components of the gas velocity are $v_{{\rm g}r}=0$,
$v_{{\rm g}z}=0$, and $v_{{\rm g}\phi} \equiv (1-\eta)\OmegaK r$. In
$z$, the particle's equation of motion reads
\begin{equation}
\ddot{z} = -\dot{z}/\ts - \OmegaK^2 z
\end{equation}
where the last term accounts for stellar gravity, for $z \ll r$.
For $\taus \ll 1$, the particle settles at terminal velocity $-\OmegaK^2z \ts$.
For $\taus \gg 1$, the particle behaves as a lightly damped harmonic oscillator
whose amplitude decays as $e^{-t/(2\ts)}$. A characteristic time for settling,
valid for all $\taus$, is
\begin{equation} \label{eq_tz}
t_z \sim \frac{1}{\OmegaK} \left( \frac{2\taus^2+1}{\taus} \right) \,.
\end{equation}
In $r$ and $\phi$,
\begin{eqnarray}
\ddot{r} - r\dot{\phi}^2 = -\vk^2/r - \dot{r}/\ts \\
r\ddot{\phi} + 2 \dot{r}\dot{\phi} = - (r\dot{\phi} - v_{{\rm g}\phi} )/\ts \,.
\end{eqnarray}
Writing $\dot{\phi} = \OmegaK +  \delta v_\phi/r$ where $|\delta v_\phi| \ll
\OmegaK r$, we approximate $\ddot{\phi} \approx \dot{\Omega}_{\rm K} \approx
-3\OmegaK \dot{r}/ (2r)$ and drop $|\ddot{r}| \ll |\dot{r}/\ts|$. Both
approximations can be checked a posteriori. To first order in
$\delta v_\phi$,
\begin{eqnarray}
\dot{r} \approx -2 \eta \OmegaK r \left( \frac{\taus}{1+\taus^2} \right) \label{eq_rdot} \\
\delta v_\phi =  (\dot{\phi} - \OmegaK) r \approx \frac{-\eta \OmegaK r}{1+\taus^2} \,. \label{eq_phidot}
\end{eqnarray}
The time for radial drift, $t_r \equiv |r/\dot{r}|$, is
$\sim$$\eta^{-1}$ longer than $t_z$. Both times are minimized, and
velocities relative to gas are maximized at $\sim$$\eta \vk$, for
marginally coupled bodies. This is the well-known problem that
boulders having sizes $\s_1 \approx m$ drift towards the star in $\min (t_r) \sim (\eta
\OmegaK)^{-1} \sim 200 (r/\AU)^{13/14} \yr$, too quickly to form
planets.

\subsubsection{Pileups and Pressure Traps}\label{sec_pile}
As particles of a given size drift inward, they tend to ``pile up,''
increasing $\Sigmap/\Sigmag$ 
(YS, \citealp{yc04}); see Figure \ref{fig:enrich}.  Their inward mass
flux decreases with decreasing $r$ if Epstein drag applies and the
outer regions are not already depleted in solids.  Idealized pile-ups
march inward on the drift timescale with increasing amplitude.  The
actual ability of particles to pile up coherently will be affected by
ongoing evolution of the particle size distribution and by any disk
turbulence. \citet{yc04} explore one prescription for turbulent
viscosity predicated on the KHI that amplifies pile-ups, but its basis
is unproven; see \S5.2.3 of \citet{garaud}.  \citet{sv96} show, by
combining the drift and turbulent diffusion of particles (see
\S\ref{sec:turb}) with the viscous evolution of $\Sigmag$, that the
solids-to-gas ratio in a disk evolves, and often increases, as it
accretes onto the star.

A special pile-up of dust could occur just inside an ice-line
\citep[e.g.][]{idalin08}. When ``dirty snowballs'' drift towards and
evaporate inside the ice-line, they unleash small dust grains that may
accumulate there \citep[see][who included this effect in a coagulation
model]{cc06}.  Furthermore, the ice-line may act as a ``cold trap'':
vapor that diffuses radially outward and crosses the line will
condense and accumulate \citep{stev88}.  Though water is the dominant
volatile, the methane condensation front could also be significant
(\S\ref{sec_disks}).

We have shown that inward particle drift is a direct consequence of
gas pressure decreasing radially outward. By the same physics, solids
of all sizes drift into and collect within local pressure maxima
\citep{whi72}, with $\taus \sim 1$ particles accumulating fastest.  If
a pressure bump has a narrow radial width $\ell < r$, even small
amplitudes $\delta P \sim (\ell /r)P$ will produce local maxima.
Proposed sources of pressure bumps include gaseous spiral arms
\citep{rice}, anticyclonic vortices \citep{chav00}, the aforementioned 
ice-lines, and fluctuations in magneto-rotational turbulence \citep[][see
\S\ref{sec:turbcon}]{jyk09}, including spatial variations in the background flux 
\citep{kato} or magnetic resistivity, as at the edges of dead zones 
\citep{gammie96}.  \citet{houches} discusses pressure trapping
in more detail.

\subsection{Turbulent Stirring} \label{sec:turb} The response of
solids to turbulent gas is multi-faceted (see \citealt{toschi},
especially their section 4.1).  Over many eddy times, turbulence
diffuses solids, vertically and radially. On shorter timescales,
transient flow structures can have just the opposite effect: they can
collect particles.

Many studies assume that turbulence in disks is Kolmogorov in
character, and in our example calculations below we also adopt this
view.  We associate $\vo$, $\to$, and $\ello \sim \vo \to$ with the
speed, turnover time, and length of the largest and fastest ``outer
scale'' eddies.  Smaller eddies of length scale $\ell$ have slower
speeds $\delta \vell \sim \vo (\ell/\ello)^{1/3}$ and shorter turnover
times $t_\ell = \ell/\delta \vell \propto \ell^{2/3}$.  The cascade of
energy from larger to smaller scales terminates at the ``inner
scale,'' characterized by $\elli \sim \nu^{3/4}\to^{1/4}/\vo^{1/2}$,
$\ti \sim \sqrt{\nu \to}/\vo$, and $\vi \sim \elli/\ti$ set by the
molecular viscosity $\nu$.  For many estimates we assume that outer
scale eddies have lifetimes limited by orbital shear, so that $\to
\sim \OmegaK^{-1}$. By ignoring the difference between the turbulent
diffusivity for mass, $\Dg \sim \vo \ello \sim \vo^2 \to$, with the
turbulent diffusivity for angular momentum, $\alpha \cg \hg \sim
\alpha \cg^2 \OmegaK^{-1}$, we extract the convenient prescriptions
$\vo \sim \sqrt{\alpha} \cg$ and $\ello \sim \sqrt{\alpha} \hg$
(\citealt{cuz01}, see also \citealt{yc04}).

However the assumption of homogeneous, isotropic turbulence characterized
by cascades of energy down to smaller scales may be incorrect.
Simulations of MRI turbulence, both global
\citep{fn05} and in large shearing boxes \citep{jyk09}, show that
inverse cascades produce pressure perturbations that survive many tens
of orbits and have in-plane sizes larger than $\hg$.  These coherent
structures may accumulate particles by pressure trapping (Figure \ref{fig:passiveconc}, \S\ref{sec_pile}).

\subsubsection{Particle-Particle Velocities} \label{sec:ppv} In active
regions of the disk, relative gas-particle and particle-particle
velocities may be dominated by turbulence \citep{kusaka, cameron},
rather than the systematic drifts considered in \S\ref{sec_lam}.
\citet{volk} derive integral expressions for these relative velocites,
evaluating them numerically for Kolmogorov turbulence.  Consider
identical grains with $\ti < \ts < \to$, i.e.\ stopping times between
the inner and outer
scales.  
Collision speeds are acquired from
eddies with a turnover time $t_\ell \sim \ts$. 
If $t_\ell \ll \ts$, coupling to the eddy is too
weak to excite particle motion.  If $t_\ell \gg \ts$, grains are so
tightly coupled to the incompressible gas flow that they do not
collide.  Thus colliding grains have relative velocity $\vpp \sim \delta \vell
\sim \vo (\ts/\to)^{1/2}$---which is not to be confused with the random speed of an individual particle, $\vp \sim \vo$. 

For $\ts > \to$, particles are loosely coupled to all eddies.  During
$\ts$ there are $\ts/\to$ kicks from the strongest, outer scale
eddies, so the random walk in particle velocity saturates at $\vp \sim
\vo/\sqrt{\ts/\to}$ \citep{yl}.  Since the random walks of different
particles are uncorrelated, the random velocity is now also the typical relative
speed.

Combining our expressions for collision speeds gives 
\begin{equation}\label{eq:vcoll}
\vpp \sim \vo \sqrt{\St/(1+\St^2)}
\end{equation}
where the Stokes number $\St \equiv t_s/t_o$.  This agrees roughly
with \citet{volk}.  \citet{weid84} provides more precise fitting
formulae---further refined by \citet{ormelcuzzi}---that include
important corrections for $\ts < \ti$ and unequal particle sizes.
\citet{yl} confirm that orbital dynamics introduces only modest 
corrections when $\to \lesssim \OmegaK^{-1}$.

Thus collision speeds in disks peak at $\vo$ for $\taus \sim 1$
particles, since $\St \sim \tau_s$ when $\to \sim \OmegaK^{-1}$.
Magneto-rotational turbulence with $\vo \sim \sqrt{\alpha} \cg \sim 70
(\alpha/10^{-2})^{1/2} \\ (r/\au)^{-3/14} \m\s^{-1}$ can induce collision
speeds faster than drift speeds (eqs.~\ref{eq_rdot} and \ref{eq_phidot}).

\subsubsection{Diffusion and Particle Scale Height}\label{sec_diffheight}
On timescales $> \to$, turbulence diffuses particles spatially.
\citet{yl} compute, formally and by order-of-magnitude methods, the
particle diffusivity $\Dp$ as a function of the gas mass diffusivity
$\Dg$.
When $\taus$ is small, orbital dynamics are negligible. If further the
Stokes number $\St \equiv \ts/\to < 1$, then particles are well
coupled to gas and $\Dp \sim \Dg$. If $\St > 1$, it takes $\ts$ for
the loosely coupled particle to have its random velocity $\vp$
(derived in the paragraph just above eq.~\ref{eq:vcoll}) changed by
order unity.  Then $\Dp \sim \vp^2 \ts \sim \vo^2 \to \sim \Dg$.  Thus
when $\taus \lesssim 1$, $\Dp \sim \Dg$ regardless of $\St$.  In
general, $\Dp \sim \Dg/(1+\taus^2/4)$ because orbital epicycles limit
diffusion \citep{yl}.

The time for particles to diffuse radially $r^2/\Dp$ is shorter than
the drift time $t_r$ when $\alpha > \taus$. Thus for $\alpha \sim 10^{-2}$, sub-cm-sized particles are
well coupled to the disk accretion flow; but see \citet{tak02} for
possible height-dependent complications.

By equating the vertical diffusion time $\hp^2 / \Dp$ with the
gravitational settling time $t_z$, we find a turbulent dust scale
height
\begin{equation}\label{eq_hp}
\hp \sim \sqrt{\frac{\Dg}{\OmegaK\taus}} \sim \sqrt{\frac{\alpha}{\taus}} \hg
\end{equation}
for all $\taus$.  \citet{carb} derive and numerically
confirm (\ref{eq_hp}) for $\taus \gg 1$.  The $\taus \ll 1$ limit is
well-known \citep{cuzzi93, dms95}, but see the end of
\S\ref{sec:turbcon} for a possible correction.

\subsubsection{Turbulent Concentration Between Eddies}\label{sec:turbcon}
On timescales $< \to$, particles tend to be centrifugally flung out of
high vorticity eddies, and funneled into their interstices of lower
vorticity \citep{max87,eatonfessler,toschi}.  Since centrifugal
support in eddy vortices makes them low pressure centers, turbulent
concentration is consistent with the tendency of particles to seek
high pressure (\S\ref{sec_pile}, \citealt{houches}).  Particles of
given $\ts$ are concentrated preferentially by eddies that turn over
on the same timescale; in other words, particles are concentrated most
strongly by eddies to which they are marginally coupled.
In a Kolmogorov cascade, smaller eddies
are more effective than larger eddies at concentrating
their respective, marginally coupled particles.
This is because vorticity $\sim \delta
\vell/\ell \propto \ell^{-2/3}$, down to the dissipation scale where
$\elli \sim
5 (r/\au)^{135/56} \m$ and $\ti
\sim 30 (r/\au)^{9/4} \s$  
for $F = 2$, $\alpha \sim 0.01$, and $\to \sim \OmegaK^{-1}$.
Turbulent concentration is special since it can collect $\taus \ll 1$
particles. Those particles that are optimally concentrated are
marginally coupled to the smallest eddies: they have $\ts = \ti$,
which corresponds to $s = \si \sim 0.1 \mm$ at 1 AU and $\si \sim 10
\mum$ at 30 AU, with some uncertainty in the viscosity (see footnote
\ref{foot_chapcow}).

\citet{cuz01} apply turbulent concentration to chondrules: objects
having sizes 0.1--1 mm are optimally concentrated at $r = 2.5$ AU in
their model disk, consistent with our estimate of $\si$ above.  The
shape of the chondrule size distribution is well reproduced by
turbulent concentration. This raises two interesting possibilities:
(1) the chondrule size distribution was initially much wider, and has
since been narrowed to the one observed today by turbulent concentration;
or less likely (2) chondrules happen to form with sizes that closely
match optimal concentration scales at the location of the asteroid belt.

The upper limit on concentration is set by mass loading, the
backreaction of particles on the turbulence.  In direct numerical
simulations of $\si$-sized particles in turbulent gas, \citet{hc07}
find a maximum concentration factor $\Phi \equiv \rhop/\rhog \approx
100$ at the grid, i.e.\ dissipation scale.  This maximum value obtains
for moderate Reynolds numbers $\Reo = \ello \vo/\nu$ and a
box-averaged $\langle \Phi \rangle = 1$.  Extrapolating these results
using ``cascade multipliers'' \citep{sreeni}, they show that $\max
\Phi \approx 100$ still applies at higher $\Reo$.  Maximal clumping at
the dissipation scale yields a mass $\Phi \rhog \elli^3$ equal to that
contained in a compact 10-cm solid at $r = 2.5$ AU. However these
clumps are unlikely to compactify because of their extremely short
characteristic lifetimes, on the order of $\ti$ \citep[][]{eatonfessler}.
For our disk parameters, $\ti \sim 4$ minutes.

\citet[][hereafter CHS08]{chs08} introduce the possibility that
$\si$-sized solids concentrate on much larger length scales $\ell_\ast
\sim 10^4 \km$, or $10^5 \elli$ at $r = 2.5\AU$.  The clump mass
$\sim$$\Phi \rhog \ell_\ast^3$ would correspond to a compact
$\sim$20-km planetesimal, for $\Phi \sim 100$ concentration.  Large
scale fluctuations might arise as an ``intermittent'' phenomenon:
deviations from self-similarity that show up in high order structure
functions \citep{frisch}.  Evidence exists for particle clumping at
larger scales, i.e. in the inertial range of isotropic turbulence, but
preferentially for particles larger than $\si$, with $\ts > \ti$
\citep{bec2007}. This is to be expected since eddies in the inertial
range have turnover times longer than $\ti$. Whether clumping of
$s_i$-sized or larger particles happens at astrophysically
interesting amplitudes and rates is an active area of research.

CHS08 find that an $\ell_\ast$-scale clump is massive enough to
survive ram pressure stripping as it plows through gas.  They
propose that self-gravity draws solids to the clump center at the
terminal velocity, a process that lasts $\sim$100 orbits.  The
turbulence that is invoked to create the clump could still destroy it,
since the lifetime of an $\ell_\ast$-eddy at 2.5 AU is $t_\ast \sim \to
(\ell_\ast/\ello)^{2/3} \sim 0.007$ orbit.  Streaming instabilities
(\S\ref{sec:SI}) may aid in clump survival, as they show that mass
loading can promote particle clumping in Keplerian disks.  The
enticing possibility that chondrules comprise first-generation
planetesimals warrants further investigation in this area.

Eq.~(\ref{eq_hp}) for the turbulent dust scale height might need revision
because turbulent concentration accelerates vertical settling
\citep{max87}.  The correction has not been investigated, but would
apply for $\St \ll 1$. Turbulent concentration is also neglected in
studies of grain growth by sticking, a subject to which we now turn.

\section{PARTICLE GROWTH BY STICKING}\label{sec:stick}
Particles stick upon colliding if they move slowly enough, dissipate
enough energy during impact, and are small enough---since surface area
increases relative to mass for smaller bodies.  We can estimate the
relevant orders of magnitude by comparing the initial kinetic energy
of two identical elastic spheres colliding at speed $v_{\rm col}$,
with the surface binding energy at the moment of maximum
deformation. From Hertz's 
law of contact, the radius of the compressed cap is $b \sim s (\rhos
v_{\rm col}^2/E)^{1/5}$, where $E$ is Young's modulus.  The binding
energy is $\sim$$\gamma b^2$, where $\gamma$ is the surface tension
from unsaturated bonds, and exceeds the collisional energy for
\begin{eqnarray} \label{eq_stick}
v_{\rm col}  < v_{\rm stick}   \sim  2.8 
\left( \frac{\gamma}{370 \erg \cm^{-2}} \right)^{5/6} \left( \frac{7 \times 10^{10} \erg \cm^{-3}}{E} \right)^{1/3}  \\
  \times \left( \frac{1 \gm \cm^{-3}}{\rhos} \right)^{1/2} \left( \frac{\mum}{s} \right)^{5/6} \m \s^{-1} \nonumber
\end{eqnarray}
for ice \citep{chokshi, youdin04}.  For $\mum$-sized spheres made of silicate, (\ref{eq_stick}) yields $v_{\rm
  stick} \sim 7 \cm \s^{-1}$, about an order of magnitude smaller than sticking velocities measured experimentally \citep{blumwurm08}.


In laminar disks, growth by sticking is expected to stall at
$\sim$cm sizes. From \Eq{eq_rdot}, radial drift velocities 
exceed $\sim$$1 \m\s^{-1}$---the sticking velocity for $\mum$-sized
monomers---once $\taus \sim 0.02$, corresponding to
particle sizes $s \sim 5 \cm$, $4 \cm$, and $2$ mm at 1, 5, and 30 AU,
respectively, for $F = 2$. Thus super-cm particles would fail to accrete
smaller grains that move with the gas.  As judged by \Eq{eq_stick},
comparably-sized, super-cm bodies would stick only at unrealistically
low relative speeds, with extremely low surface binding energies per unit
mass. Indeed bodies $\gtrsim$ mm in size have not
been observed to stick experimentally \citep{blumwurm08}.

Particle porosity abets sticking by allowing for greater dissipation
of kinetic energy \citep{dt97}. Moreover porosity lowers $\rhos$,
which permits larger masses to coagulate when growth is limited to a
fixed $\taus$, as above. The degree of enhanced mass growth depends on the drag
regime; fixing the stopping time gives a particle mass $m
\propto \rhos^{-2}$ for Epstein drag, and $m \propto
1/\sqrt{\rhos}$ for Stokes drag.  These advantages are limited,
however, since collisions at speeds of $0.1$--$1 \m \s^{-1}$ restructure and
compactify aggregates composed of $\mum$-sized monomers \citep{dt97}.

Simulations of grain growth by \citet{dd05} and \citet{ormel} include
vertical settling and neglect radial drifts. \citet{dd05} find that in
laminar disks, porous aggregates achieve equivalent compact sizes of a
few cm before settling to the midplane, while particles assumed to be
always compact grow up to 1 cm (see their Figure 3, models S2, S5, and
S6).  \citet{ormel} calculate that in turbulent disks, porous bodies
attain equivalent compact sizes of several cm, for $\alpha =
10^{-2}$ at $r = 1$ and 5 AU.

In both these studies, $\mum$-sized grains deplete, and the disk
becomes optically thin, within $\sim$$10^3$ yr. This timescale is too
short to be reconciled with astronomical observations of disk spectra.
To maintain the population of small grains and the disk's optical
depth over 1--10 Myr, models may need to account for collisional
fragmentation of aggregates, or condensation of dust from silicate
vapor in hot, active disk regions \citep{dd05,ormel}.  Turbulent
concentration of dust (\S\ref{sec:turbcon}) could also be an important
correction to coagulation models, not only because densities are
enhanced, but because collision speeds are likely
reduced in particle clumps, as measured in 
simulations of the streaming instability \citep[][ \S\ref{sec:SI}]{jym}.

\section{CRITERIA FOR GRAVITATIONAL INSTABILITY}\label{sec:criteria}
\subsection{Dynamical Collapse}\label{sec:dyncollapse}
We analyze the stability of a planar, self-gravitating, rotating sheet
of dust, modeled as a fluid (GW, \citealt{saf}). For the moment we ignore
interactions with gas. Unperturbed, the
dust has surface density $\Sigma$, angular speed $\Omega$, and
barotropic pressure $P(\Sigma) \sim \Sigma c^2$, where $c$ is the
velocity dispersion. For axisymmetric
perturbations, the linearized equations for continuity and momentum
read \citep[e.g., chapter 6 of][]{bt}
\begin{eqnarray}
\frac{1}{\Sigma} \frac{\partial \Sigma'}{\partial t} + \frac{\partial v_r'}{\partial r} = 0 \label{eq_pert1}\\
\frac{\partial  v_r'}{\partial t} - 2 \Omega v_{\phi}' = \frac{-c^2}{\Sigma} \frac{\partial \Sigma'}{\partial r} - \frac{\partial \Phi'}{\partial r} \label{eq_pert2} \\
\frac{\partial v_{\phi}'}{\partial t} + \frac{\kappa^2}{2\Omega} v_r' = 0 \label{eq_pert3}
\end{eqnarray}
where perturbations are primed, $\Phi'$
is the perturbation gravitational potential, and $\kappa = (r
d\Omega^2/dr + 4 \Omega^2)^{1/2}$ is the epicyclic frequency of radial
oscillations---in practice nearly equal to $\Omega$, its value for a
Kepler potential. For WKB waves, perturbations $\propto e^{\imath(k_rr + \omega t)}$
where $k_rr \gg 1$, and $\Phi' \approx -2\pi G\Sigma' / |k_r|$
\citep{bt}. Then (\ref{eq_pert1})--(\ref{eq_pert3}) yield the well-known
dispersion relation for axisymmetric waves:
\begin{equation}
\omega^2 = c^2k_r^2 - 2\pi G \Sigma |k_r| + \kappa^2
\end{equation}
which informs us that pressure stabilizes short wavelengths, rotation
stabilizes long ones, and self-gravity de-stabilizes intermediate
wavelengths when
\begin{equation} \label{eq_Q}
Q \equiv \frac{c\kappa}{\pi G \Sigma} < 1
\end{equation}
\citep{toomre64,glb}. Modes whose wavelengths
exceed $\lambda_{\rm crit} = 4\pi^2 G\Sigma/\kappa^2$ (neutrally stable
for $c=0$) are always stable. The fastest growing mode has
$\lambda_{\rm fgm} = 2c^2 / (G\Sigma) = Q^2 \lambda_{\rm crit}/2$.

A standard expression for the vertical thickness of the dust layer
is $\hp \approx c/\Omega$ (caveat emptor: (\ref{eq_hp}) shows
this is not valid when $\St \ll 1$). Then Toomre's criterion
(\ref{eq_Q}) translates into a thickness criterion
\begin{equation}\label{eq_hpstar}
\hp < \hp^{\ast} \approx \frac{\pi G \Sigmap}{\Omega^2} \approx 2 \times 10^8 F \Zr \left( \frac{r}{\au} \right)^{3/2} \cm 
\end{equation}
or equivalently, a requirement on the midplane density
\begin{equation}
\rho > \rho^{\ast} \approx \frac{\Sigma}{2\hp^{\ast}} \approx \frac{M_{\ast}}{2\pi r^3} \approx 10^{-7} \left( \frac{r}{\au} \right)^{-3} \gm \cm^{-3} \,,
\end{equation}
which has the same scaling as the Roche criterion (\S\ref{sec_intro}),
but is smaller by a geometric factor of $\sim$20 that arises because
the gravity of an axisymmetric ring is greater than that of a
tidally distorted ellipsoid.
The conditions for dynamical collapse differ
by orders of magnitude from those afforded by the gas disk:
\begin{eqnarray} \label{eq:hpstarhg}
\hp^\ast/\hg &\approx& 5 \times 10^{-4} F \Zr \left( \frac{r}{\au} \right)^{3/14} \\
\rho^\ast/\rhog &\approx& \frac{35}{F} \left( \frac{r}{\au}\right)^{-3/14} \,.
\end{eqnarray}
Turbulence would have to be exceptionally weak to allow particles
to sediment to such a thin layer: from
(\ref{eq_hp}), $\alpha$ would need to be $\lesssim 3 \times 10^{-7}
\taus (F \Zr)^2 (r/ \au)^{3/14}$. Overcoming this
obstacle---and showing that there are alternatives to dynamical
collapse from a disk of uniform Solar abundance---are the goals of this
review, with some of the main ideas summarized in Figures
\ref{fig:enrich} and \ref{fig:ladder}.

\citet{sek83} solves for the linear stability of a
midplane layer composed of perfectly coupled dust and gas.  The layer is
confined by the pressure of overlying gas layers.
As with Toomre's calculation above, axisymmetry is assumed; however the midplane
layer's vertical thickness is free to change.  
Dynamical instability occurs for midplane densities about 4 times greater
than $\rho^{\ast}$ because gas pressure helps to stabilize the midplane layer.
Since the layer is incompressible, in-plane motions give rise to
out-of-plane ``bulges.'' Though not modeled, slow sedimentation of dust to
the centers of these bulges could produce km-scale planetesimals.
 
Toomre's criterion (\ref{eq_Q}) is a good rule of thumb for deciding
when self-gravity matters, but it misleads because it is derived for
axisymmetric waves.  In reality, GI proceeds non-axisymmetrically.
Non-axisymmetric waves amplify when swung from leading to trailing by
the background radial shear \citep{glb}. Waves grow even if $Q > 1$,
but especially strongly as $Q \rightarrow 1$.  \citet{t81} gives a
delightful tutorial, explaining swing amplification as a
near-resonance between radial shear, epicyclic motion, and
self-gravity.  Amplification is restricted to the interval, of
duration $\sim$$1/\Omega$, when wave pitch angles slew from about -1
to 1 rad; upon completion, $\lambda_r \sim \lambda_\phi$ and
amplitudes have grown by factors $\gtrsim 50$ for $Q \lesssim 1.2$.
Modes most prone to growth have $\lambda_\phi \approx \lambda_{\rm
  crit}$. For these reasons, GW describe the disk's initial fragments
as having a characteristic length $L_{\rm frag} \sim \xi \lambda_{\rm
  crit}$ and mass $M_{\rm frag} \sim \Sigma \xi^2 \lambda_{\rm crit}^2
\equiv \xi^2 M_{\rm crit}$, where $\xi$ is an order unity parameter
that contains our uncertainty about the spectrum of seed perturbations
and how close $Q$ is to unity. At $r = 1\AU$, $M_{\rm frag}$ equates
to a compact rocky planetesimal having size $s_{\rm frag} \approx 8 (F \Zr/0.33) \xi^{2/3} \km$; this estimate figures prominently in the lore of
``kilometer-sized planetesimals.''

Numerical N-body simulations of inelastically colliding particles
verify that within 1--2 orbital periods, particles aggregate on the
scale $\lambda_{\rm crit}$ \citep[]{michi}.  The linear mass
scale $M_{\rm crit}$ is not as prominent in either particle or gas
simulations \citep{gam01}, partly because fragments 
accrete rapidly---suggesting that $M_{\rm crit}$ be used cautiously, and
probably as a lower limit.

\subsection{Drag-Assisted Gravitational Instability}\label{sec:draggi}
The stability properties discussed above pertain to dust treated
as a single frictionless fluid. But gas-dust interactions can change
this picture qualitatively. The simplest modification is to introduce
drag terms to the momentum equations \citep{ward76,ward00,coradini,youd05a}:
\begin{eqnarray}
\frac{\partial v_r'}{\partial t} - 2 \Omega v_{\phi}' = \frac{-c^2}{\Sigma} \frac{\partial \Sigma'}{\partial r} - \frac{\partial \Phi'}{\partial r} - \frac{ v_r'}{\ts} \label{eq_pert2a} \\
\frac{\partial v_{\phi}'}{\partial t} + \frac{\kappa^2}{2\Omega} v_r' = - \frac{v_\phi'}{\ts} \,.\label{eq_pert3a}
\end{eqnarray}
The meaning of $v'$ bears clarification. With drag,
particles drift radially (\S\ref{sec_lam}). Therefore perturbed
quantities refer to a background with steady $v_r \neq 0$, and our WKB
analysis is restricted to $t < t_r$. Backreaction on gas, whose
properties are assumed fixed, is neglected.

When $\taus \ll 1$, the drag-modified
equations yield a new dispersion relation:
\begin{equation}
\omega = \imath (c^2 k_r^2 - 2\pi G \Sigma |k_r|)\ts
\end{equation}
which implies that modes for which $k_r < 2\pi G \Sigma / c^2$ are
unstable---even when $Q > 1$! The fastest growing mode has
$\lambda_{\rm fgm} = 2 c^2 / (G\Sigma) = Q^2 \lambda_{\rm crit}/2$ and
growth rate $|\omega_{\rm fgm}| = \Omega \taus / Q^2$. Thus drag
destabilizes long wavelength modes, on longer timescales, compared to
the dissipationless case. These qualitative differences---in
particular the removal of Toomre's $Q$-criterion---apply for a variety
of drag regimes, and for cases that include turbulent diffusion of
particles, accounted for by adding $(\Dp/\Sigma) \partial^2 \Sigma'
/ \partial t^2$ to the right-hand side of (\ref{eq_pert1}) (Youdin, in
preparation; Karim Shariff \& Jeffrey Cuzzi, personal communication).

\citet[hereafter GP00]{gp00} explain the instability simply. In the
unperturbed state, dust has constant surface density and drifts
radially inward at constant velocity. An axisymmetric overdense ring
of width $\Delta r$ exerts a gravitational pull radially inward at its
outer edge. Provided $\hp \ll \Delta r \ll \hg$, dust rotates faster
there, while the gas velocity is unaltered. The increased drag causes
dust to flow at a greater rate into the annulus. Likewise there is a
flow into the annulus at the inner edge, where dust experiences less
drag.  Thus when combined with gas drag, self-gravity, no matter how
weak, can draw particles into rings.

The stability of non-axisymmetric modes with dissipation is a largely
open field. \citet{noh} make some exploratory integrations of the
linearized perturbation equations.  In N-body simulations that include
$-v' / \ts$ drag and a fixed gas velocity field, particles cluster
more readily because gas drag damps their velocity dispersion
\citep{tanga}.

\section{THIN DUST LAYERS: VERTICAL SHEARING \\
  INSTABILITIES}\label{sec:thin}

Particles may be prevented from settling to the midplane not only by
turbulence intrinsic to the gas (e.g., sustained by the MRI), but also by
turbulence triggered by the particles themselves \citep{weiden80}.  We
consider here instabilities caused by small $\taus \ll 1$ solids that
drive a vertical shear.  In \S\ref{sec:DI} we show that larger
particles, in addition to inducing a vertical shear, can drive
turbulence of a different character via relative streaming motions.


As dust settles, a particle-rich sublayer develops which orbits at
nearly the full Keplerian velocity. The dust-poor gas above and below
shears by at speeds approaching the pressure-supported value $\eta
\vk$ (eq.~\ref{eq_etavK}).  Such a stratified shear flow is subject to
a Kelvin-Helmholtz-type instability (KHI).  For a non-rotating
flow---which we later show turns out to be a fair analogue for actual
vertically shearing flows in disks---the KHI can develop when the
Richardson number
\begin{equation} \label{eq_ri}
\Ri = \frac{ (g/\rho) \partial \rho / \partial z}{(\partial v_{\phi}/\partial z)^2} < \frac{1}{4}
\end{equation}
somewhere \citep{drazin}, where $g$ is the vertical
gravitational acceleration and $\rho = \rhop+\rhog$ \citep[see][for the applicability of $\Ri$ to a dust-gas mixture]{garaud}.  The square root
of $\Ri$ is the Brunt-Vaisala frequency for buoyant vertical
oscillations, divided by the shear rate $\partial v_{\phi}/\partial
z$. This is a sensible criterion: when the buoyancy frequency is
small, vertically displaced fluid elements fail to return to their
equilibrium positions before being whisked away by the background
shear.

When we neglect self-gravity and approximate $g \sim -\OmegaK^2\hp$,
and further take $\partial (\ln \rho) / \partial z \sim -1/\hp$ and
$|\partial v_{\phi}/\partial z| \sim \eta \vk / \hp$, criterion
(\ref{eq_ri}) implies that the KHI prevents the sublayer from having a
thickness less than
\begin{equation}\label{eq_hpRi} 
h_{\rm p,Ri} \sim \Ri^{1/2} \eta r \sim 10^{-2} \left( \frac{r}{{\rm AU}} \right) ^{2/7} \hg \, .
\end{equation}
Thus the midplane density falls short of $\rho^{\ast}$ by a factor of
$\sim$$20/ (F \Zr)$, roughly consistent with findings by
\citet{cuzzi93}, who use mixing length models for particle-laden
turbulence scaled to laboratory experiments.  If we insist that the
disk be of bulk Solar composition, then achieving higher densities
requires either that particles be larger ($\taus \gtrsim 1$) so as to decouple
from the turbulence, or that they clump locally (\S\ref{sec:turbcon},
\S\ref{sec:DI}).

Nominally we can estimate, using (\ref{eq_hp}), that turbulence
intrinsic to gas dominates the KHI when $\alpha \gtrsim \taus 
(h_{\rm  p, Ri}/\hg)^2 \sim 3 \times 10^{-4} \taus (r/\au)^{4/7}$.
Whether MRI-dead zones (\S\ref{sec_disks}) are sufficiently passive
for the KHI to manifest remains an outstanding question.
Our $\alpha$ scalings may not apply to the weak motions
possibly excited by MRI-active surface layers
(\citealt{gammie96, fs03, oishi07}; see also \citealt{bai} for detailed estimates of the active layer thickness).

In one of the earliest works to treat the KHI in disks realistically
by accounting for orbital differential rotation, \citet{is}
numerically integrate the linearized perturbation equations and find
that maximum growth factors increase strongly with midplane
dust-to-gas ratios---so strongly that $\rhop/\rhog$ is likely to stall
at values just above unity in disks having bulk solar metallicity.
Their conclusion is borne out by shearing box simulations
\citep{c08,b09}, performed in the limit that dust is perfectly coupled
to gas ($\taus \rightarrow 0$). These simulations demonstrate that
despite the inclusion of orbital dynamics, Richardson numbers
characterizing instability remain between 0.1 and 1 (Figure
\ref{fig:khi}).  In retrospect this is not surprising, since
rotational and radial shearing frequencies $\sim$$\OmegaK$ are only of
the same order as both the Brunt-Vaisala frequency, and the vertical
shearing frequency $h_{\rm p,Ri}/(\eta \vk) \sim
\Ri^{1/2}\OmegaK^{-1}$, when $\Ri \sim 1$. Nevertheless radial shear
is critical for limiting the growth of perturbations that otherwise
run away when only the Coriolis force is included \citep[cf.][]{go}.

As pointed out by J.~Goodman (personal communication; see
\citealt{c08}), the KHI may not be the only instability at play. Dusty
sublayers may also be baroclinically unstable: their isodensity
surfaces, dominated by dust, do not align with their isobars, dictated
by gas \citep{knob}.  Axisymmetric baroclinic stability is assured by
\citet{is} and by analogy with the Solberg-Hoiland criteria for
rotating stars \citep{kip}.  Specifically, displacements at constant
entropy are almost perfectly radial when vertical stratification of
the dust is strong.  Thus in a Keplerian disk, outward adiabatic
displacements move toward regions of higher specific angular momentum,
and are Rayleigh stable.

Aside from postulating large, aerodynamically decoupled particles,
another way to prevent the shearing instability from forestalling GI
is to substantially raise $\Zr$. \citet{sek98} discovers that for
flows with constant $\Ri(z) = 1/4$, $\Zr \approx 6$--30 can lead to
$\rho(z=0) \approx \rho^{\ast}$ in a midplane cusp.  In fact, by
further increasing $\Zr$ slightly, and accounting properly for
vertical self-gravity, the cusp becomes singular: $\rho(0) \rightarrow
\infty$.  YS perform similar calculations, finding $\rho \approx
\rho^{\ast}$ could be achieved for $\Zr$ as low as 2, depending on
disk parameters.  They show that the cusp develops when $\rhop \gtrsim
\rhog$ throughout the layer, which from (\ref{eq_hpRi}) requires
$\Sigmap/\Sigmag \gtrsim \sqrt{\Ri} \eta \vk/\cg$, or $\Zr\gtrsim 1
(r/{\rm AU})^{2/7}$.  Physically, as solids are added, the buoyancy
frequency increases, while the shear saturates at $\eta \vk/\hp$.
Numerical simulations, which as yet do not include vertical
self-gravity, suggest $\Zr \gtrsim 5$ to attain densities approaching
$\rho^{\ast}$ \citep{c08}.  Mechanisms to achieve supersolar
metallicities are shown in Figure \ref{fig:enrich} and described in
\S\ref{sec_pile}.


Whether or not they are gravitationally unstable,
thin dust layers for which $\rhop \gtrsim \rhog$ are also prone
to the streaming instability (\S\ref{sec:SI}), a mechanism
that can further concentrate solids relative to gas when $\taus \neq 0$.

\section{DRAG INSTABILITIES} \label{sec:DI} The vertical shearing
instability described in \S\ref{sec:thin} arises in dense particle
layers, even when dust is perfectly coupled to gas ($\taus \rightarrow
0$).
When relative motion between gas and dust is allowed ($\taus \neq 0$),
new secular (i.e.~requiring dissipation) instabilities arise.  These drag
instabilities can strongly clump particles relative to gas.  As with
vertical shear, drag instabilities require that $\rhop/\rhog$ be large
enough that particles backreact significantly on gas.  Self-gravity is
not required, in contradistinction to the drag-assisted GI of
\S\ref{sec:draggi}.

We describe two different models of drag instabilities.  The secular
dust layer instability (SDLI) of GP00 considers the drag arising from
turbulent stresses on the stratified dust layer.  The
streaming instability (SI) of \citet[hereafter YG05]{yg05} accounts for the 
 drag forces acting mutually between gas and particles
in a laminar unstratified disk.  Both analytic models---as well as numerical
simulations that add extra physics to the SI \citep{johansen07}---show that particles
clump.

\subsection{Secular Instability of the Dust Layer}
The SDLI uses a single-fluid, height-integrated, axisymmetric model of
the dust-rich sublayer.  Dust-poor gas at the top and bottom of the
sublayer shears by at relative speed $\eta \vk$, and is envisioned to
be turbulent (\S\ref{sec:thin}). These turbulent boundary layers exert
drag on the dust layer. Turbulent surface stresses are assumed to be
communicated throughout the entire layer; the detailed vertical
response is, by construction, ignored. To calculate the turbulent
drag, GP00 adopt prescriptions from GW and \citet{cuzzi93} for
``plate drag,'' drawing an analogy with Ekman flow past a rigid
plate. The SDLI generates overdense rings on orbital timescales, with
radial widths comparable to the sublayer vertical thickness.

The use of the plate drag formula is not
well founded, except perhaps if particles in the sublayer all had
$\taus \sim 1$ \citep{yc04}. But
GP00 argue that their results are more general, transcending the
specific drag prescription used to obtain them.  The crucial
assumption underlying the SDLI is that drag is collective, i.e.~it
depends on the surface density of the sublayer.  GP00
construct a toy model that demonstrates how any collective drag force
leads to instability.  The insights apply to drag instabilities
generally, including the SI.

In the toy model, the sublayer is described by a 1D distribution of mass density
$\Sigma$ and velocity $v$:
\begin{eqnarray}  \label{eq:toy1}
{\p \Sigma \over \p t} + {\p \over \p x} \left(\Sigma v \right) &=& 0 \\
{\p v \over \p t} +v {\p v \over \p x}&=& g - \nu_{\rm d}(\Sigma) v
\label{eq:toy2}
\end{eqnarray} 
where $\nu_{\rm d}(\Sigma) > 0$ accounts for drag, and $g < 0$
serves as a proxy for gravity, pressure, and Coriolis
forces.  The equilibrium state has uniform mass density
$\Sigma_0$ and inward drift velocity $v_0 = g/\nu_{\rm
  d}(\Sigma_0) \equiv g/\nu_0$.
We might imagine $\nu_{\rm d}$ increases with $\Sigma$---insofar as the
plate drag torque is independent of $\Sigma$ while the angular
momentum of the sublayer is proportional to $\Sigma$, so that
the drift speed decreases with increasing $\Sigma$---but the details are
unimportant: instability arises as long as
$\nu_{\rm d}$ and $g$ depend on $\Sigma$ differently. With no loss of
generality, all of the $\Sigma$-dependence is relegated to $\nu_{\rm
  d}$.

We take linear perturbations, denoted by primes, to have a Fourier dependence $\exp(\Gamma t - \imath k x)$.  The drag coefficient is expanded
\begin{equation}  \label{eq:toyexpand}
 \nu_{\rm d}(\Sigma) = \nu_{\rm d}(\Sigma_0) + \left. d \nu_{\rm d}/d\Sigma\right|_0 \Sigma'\, .
\end{equation} 
A non-zero drift $v_0$ is necessary for perturbations to $\nu_{\rm d}$
to enter linearly. The dispersion relation reads
\begin{equation} \label{eq:toygrow}
\Gamma  = {\nu_0 \over 2} \left(-1+{2\imath \tilde{k} } \pm \sqrt{1 - 4 \imath \delta_{\nu} \tilde{k} } \right) \, ,
\end{equation} 
where $\tilde{k} \equiv k v_0/\nu_0$ and $\delta_\nu \equiv d \ln
\left. \nu_{\rm d}/d\ln \Sigma \right|_0$.
For any $\delta_\nu \neq 0$ the positive root gives growth, since
$\Re\left(\sqrt{1 + \imath b}\right) > 1$ for all real $b$.  Thus any
collective drag produces instability.

To understand the growth mechanism, we examine the eigenfunctions
$\Sigma'$ and $v'$, first Taylor expanding the growing mode about
$\delta_\nu$,
\begin{equation} 
\Gamma = \Gamma_0 + \Gamma_1 + \Gamma_2 = 
\imath k v_0 - \imath \delta_\nu k v_0 +  \nu_0 \tilde{k}^2 \delta_\nu^2\, ,
\end{equation} 
and similarly expanding $v' = v_0' + v_1' + v_2'$. By substituting
these into the linearized eqs.~(\ref{eq:toy1})--(\ref{eq:toy2}) and
equating terms of the same order, we see that $v_0' = 0$, $v_1' =
-\delta_\nu (\Sigma'/\Sigma_0) v_0$, and $v_2' = - (\Gamma_1 /\nu_{\rm
  d} ) v_1'= -\imath \tilde{k} \delta_\nu^2 (\Sigma' /\Sigma_0)
v_0$. For $\delta_\nu = 0$, oscillations are stable with wave speed
$\Im(\Gamma_0)/k = v_0$.  This is a neutral mode where $\Sigma'$
simply advects with the background flow. To first order in
$\delta_\nu$, $v_1'$ is anti-phased with $\Sigma'$; if $\delta_\nu > 0$,
the drift speed decreases (increases) at density maxima (minima) due
to the perturbed drag force; the wave does not amplify but its speed
shifts by $\Im(\Gamma_1)/k = -\delta_\nu v_0$.  To second order in
$\delta_\nu$, the acceleration of $v_1'$ is subject to drag, which
induces a secondary flow $v_2'$. Since $v_2'$ is $-\pi/2$ out of phase
with $\Sigma'$, the mode grows, at rate $\Gamma_2$. The phase shift
enters because accelerations change sign at density extrema.
Figure \ref{fig:DI} depicts the eigenfunctions.


The toy model shows that essential components of drag instabilities
include the background drift $v_0$, which allows collective effects to
enter linearly, and time-dependent oscillations that drag forces
overstabilize.
  
\subsection{Streaming Instability}\label{sec:SI}
Since the streaming instability involves a minimal amount of physics---Keplerian
orbital motion, gas pressure, and a drag acceleration that is linear
in relative velocity---it supports the robustness of
drag instabilities. YG05 uncover the SI by modeling
gas and dust as two interacting fluids that obey
\begin{eqnarray} \label{eq:si1}
 {D_{\rm p} \vc{v}_{\rm p} \over D t} &=& -\OmegaK^2\vc{r}  - {\vc{v}_{\rm p} - \vc{v}_{\rm g} \over \ts}  \\ \label{eq:si2}
 {D_{\rm g} \vc{v}_{\rm g} \over Dt} &=& - \OmegaK^2\vc{r}  +{\rho_{\rm p} \over \rho_{\rm g}} {\vc{v}_{\rm p} - \vc{v}_{\rm g} \over \ts} - {\nabla P \over \rho_{\rm g}} \\ \label{eq:si3}
 {D_{\rm p} \rho_{\rm p} \over D t}  &=& - \rho_{\rm p} \nabla\cdot\vc{v}_{\rm p} \\
   \nabla\cdot\vc{v}_{\rm g} &=& 0 \label{eq:si4}
 \end{eqnarray}
 where $D_i / Dt \equiv \p/\p t + \vc{v}_i\cdot \nabla$.  Vertical
 gravity is neglected, so $r$ is the cylindrical radius.
 \citet[hereafter YJ07]{yj07} confirm that gas compressibility
 is justifiably neglected in (\ref{eq:si4}), and that
 the fluid approximation for dust holds.\footnote{We are not aware
   of a formal criterion for deciding when dust can be modeled as a fluid. The
   usual criteria for gas molecules, that they be collisional,
do not apply to dust particles
entrained in gas. YG05 suggest that coupling to the gas may suffice: if $\omega \ts \ll 1$, then the fluid approximation
is expected to be valid on timescales $\omega^{-1}$.  A more rigorous criterion would be useful since many nonlinear simulations \citep[e.g.][]{hc07} use the two-fluid approach.} 


The steady-state solutions to (\ref{eq:si1})--(\ref{eq:si4}) give
Keplerian motion, plus radial and azimuthal drifts between dust and gas
\citep[][ \S\ref{sec_lam}]{nakagawa}.  The relative streaming motion
is linearly overstable (YG05), behaving similarly
to the GP00 toy model. The main complication is that while unstable modes
can be axisymmetric, they necessarily involve motions in all three
directions.

Particle clumping occurs through the backreaction term in
(\ref{eq:si2}): $-\rhop(\vc{v}_{\rm p} - \vc{v}_{\rm g})/(\rhog \ts)$.
Since $\rhog$ and $\ts$ are constant---by assumption in YG05, and
to excellent approximation for subsonic flows---only perturbations to $\rhop$
couple linearly to background streaming motions $\vc{v}_{\rm p} -
\vc{v}_{\rm g}$.  Thus particle clumping is required to extract energy
from streaming motions, as in the GP00 toy model.  The SI is
ultimately powered by the background gas pressure gradient, which does
work on radially flowing gas: $-v_{{\rm g}r} \partial P / \partial r >
0$.  Since streaming motions transport angular momentum inward, no
energy is extracted from the Keplerian shear flow (YJ07).

The SI is controlled by two parameters: $\taus$ and
$\langle\rhop/\rhog\rangle_0$, the background dust-to-gas ratio.  While
the SI always exists, growth is slow in the test-particle
limit $\rhopg \ll 1$, and in the perfect coupling limit $\taus \ll 1$.
As $\taus \rightarrow 1$, background drift speeds peak, and modes
can grow on orbital times. 

The nonlinear consequences of the SI are best studied by computer
simulations, which can selectively include other physics.  Treating
particles as a continuous fluid simplifies analytic calculations, but
causes difficulties in numerical simulations.  The use of artificial viscosity
to avoid density discontinuities in the pressureless fluid
can underestimate particle concentrations and otherwise
compromise results.  Alternatively, ``hybrid'' simulations model gas as a
fluid on an Eulerian grid and solids with Lagrangian
superparticles, each representing a swarm of actual particles.
YJ07 simulate the linear growth of SI, and confirm that the
hybrid and two-fluid approaches give convergent results.  They
also provide the eigenfunctions of growing
Fourier modes, useful for testing codes with two-way drag forces
\citep{bal08, ish09}.

\citet{jy07} perform nonlinear 3D simulations of the idealized SI
using superparticles.  Marginally coupled, $\tau_{\rm s} = 1$ solids
concentrate by factors of several hundred for a range of $0.2 < \rhopg
< 3.0$.  Better coupled $\taus = 0.1$ particles give particle
overdensities of several tens, but only for $\rhopg > 1$.  When
$\rhopg = 0.2$, particle clumping is of order unity.
Sample simulations are displayed in Figure \ref{fig:7sisters}.

Realistically assessing how effectively the SI can concentrate
particles requires that vertical stratification be included
\citep[e.g.][hereafter JYM09]{jhk06,johansen07,jym}, since high
dust-to-gas ratios also trigger the KHI (\S\ref{sec:thin}).  JYM09
perform 3D hybrid simulations of $\taus = 0.1$--$0.4$ solids to study
the combined effects of KHI and SI.  They find a metallicity threshold
for strong clumping: when $\Zr$ increases from 2/3 to 4/3, the SI,
unaided by self-gravity, triggers overdensities having $\rhop > 10^3
\rhog$. The $\Zr$ threshold corresponds to $\rhopg \gtrsim 1$ at the
midplane, similar to the analytic criterion for the saturation of the
KHI (YS).  It remains to be seen if this threshold will shift, or be
manifested differently, for smaller particles.  Hybrid simulations
with SI that include self-gravity are discussed in
\S\ref{sec_nonlinear}.

\section{GRAVITATIONAL COLLAPSE INTO THE NONLINEAR
  REGIME} \label{sec_nonlinear} The fragmentation of the disk into
marginally self-gravitating clumps is only the beginning of the story
of planetesimal formation.  About six orders of magnitude need
to be traversed from the Roche density to solid density. Even if
clumps survive buffeting by turbulence and ram pressure stripping
\citep{chs08}, their further contraction is resisted by internal
random motions and net angular momentum.  Gas drag slows the
sedimentation of particles toward clump centers, but it also enables
such settling by braking the clump's rotation and
damping random motions. For $\taus \ll 1$, collapse proceeds on
timescales longer than the free-fall time; how much longer is unclear.
At large enough densities, inelastic collisions
between particles further reduce random motions and accelerate
gravitational collapse.

The popular notion that GI produces
planetesimals having sizes on the order of a kilometer probably
originates from GW, who recognize that a fragment having the
full mass $M_{\rm frag}$ (\S\ref{sec:criteria}) has too much angular
momentum to collapse unimpeded to solid density. They propose instead
that first-generation planetesimals form from overdense regions having
horizontal sizes $\Sigmap r^{9/4}/(\rhos^{1/4} M_{\ast}^{3/4}) \ll
\lambda_{\rm crit}$; such clumps, if they conserve angular momentum,
collapse into solid objects having sizes $S \sim 0.4 F \Zr \km$,
spinning just below break-up \citep[see also \S15.2 of][]{g04}.  But
this estimate is unjustified because it assumes that scales $\ll
\lambda_{\rm crit}$ can collapse; this would require $Q \ll 1$, and
indeed GW assume unrealistically that $c = 0$.

While reliable analytic estimates of the sizes of first-generation
planetesimals are lacking, numerical simulations have made great
strides, providing a proof of principle that gravitational collapse
can occur, even in turbulent flows. \citet{johansen07} synthesize
practically all of the physics described in this review by executing
3D, self-gravitating, vertically stratified, shearing box simulations
of superparticles in MRI-turbulent gas. Particles all have $\taus =
0.25$--1.  Three ways of concentrating particles manifest: trapping
within transient maxima in gas pressure (\S\ref{sec_pile}); the
streaming instability (\S\ref{sec:SI}); and self-gravity.  Within only
a few orbital periods, the largest bound cluster of superparticles
attains a mass $\sim$$3 \times$ that of the 450-km-radius asteroid
Ceres---though collapse to solid densities is not explicitly followed
because of finite grid resolution.  Surprisingly, magneto-rotational
turbulence hastens gravitational collapse, through
pressure trapping.  The main uncertainties of these simulations are 
the assumption that sticking produces the largish seed particles, 
having sizes of 15--60 cm at 5 AU; and the specific realization of 
MRI turbulence, which depends on the unknown net magnetic flux 
and on microscopic dissipation parameters that are too small to 
simulate exactly \citep{lesur,fromang09,davis}.

MRI turbulence, though it appears to aid gravitational collapse, 
is not a necessary ingredient. Figure \ref{fig:7sisters} displays
how bound clumps also form in an unmagnetized simulation
(JYM09).

\section{SUMMARY AND OUTLOOK}
For a collection of grains within a protoplanetary disk to become
gravitationally bound, their density $\rhop$ must exceed the Roche
density, which for typical parameters is over 1000 times the ambient gas density
$\rhog$. Equivalently, the dust-to-gas ratio must be
enhanced over the protosolar value by a factor of 60,000.  Daunting as
it may seem, the requirement may be met by several mechanisms, working
in concert, that concentrate particles relative to gas.

\begin{figure}
\centerline{\psfig{figure=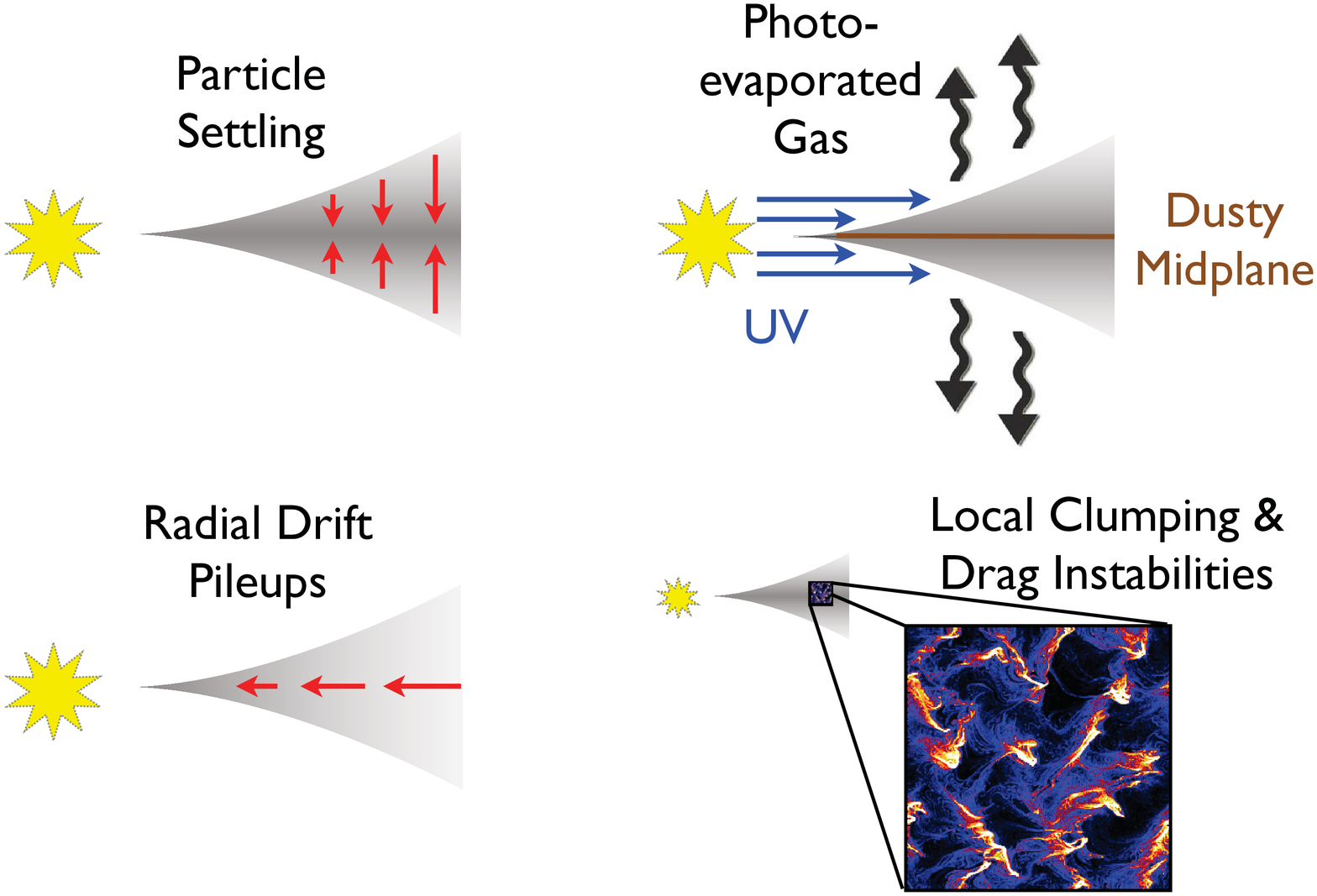,height=30pc}}
\caption{Four mechanisms for metal enrichment---none of which involve
  self-gravity. Dust can settle vertically into a dense sublayer
  \citep[e.g.,][ \S\ref{sec:thin}]{c08,b09}; pile up as it drifts
  radially \citep[][ \S\ref{sec_pile}]{ys,yc04}; remain behind as
  stellar ultraviolet radiation photoevaporates gas
  \citep[e.g.,][]{tb05}; and be concentrated on small scales by 
  passively responding to turbulent fluctuations (\S\ref{sec_pile}, 
  \S\ref{sec:turbcon}) and by actively driving drag instabilities with
  gas \citep[e.g.,][ \S\ref{sec:DI}]{gp00,yg05,jy07}.  }
\label{fig:enrich}
\end{figure}

\begin{figure}
\centerline{\psfig{figure=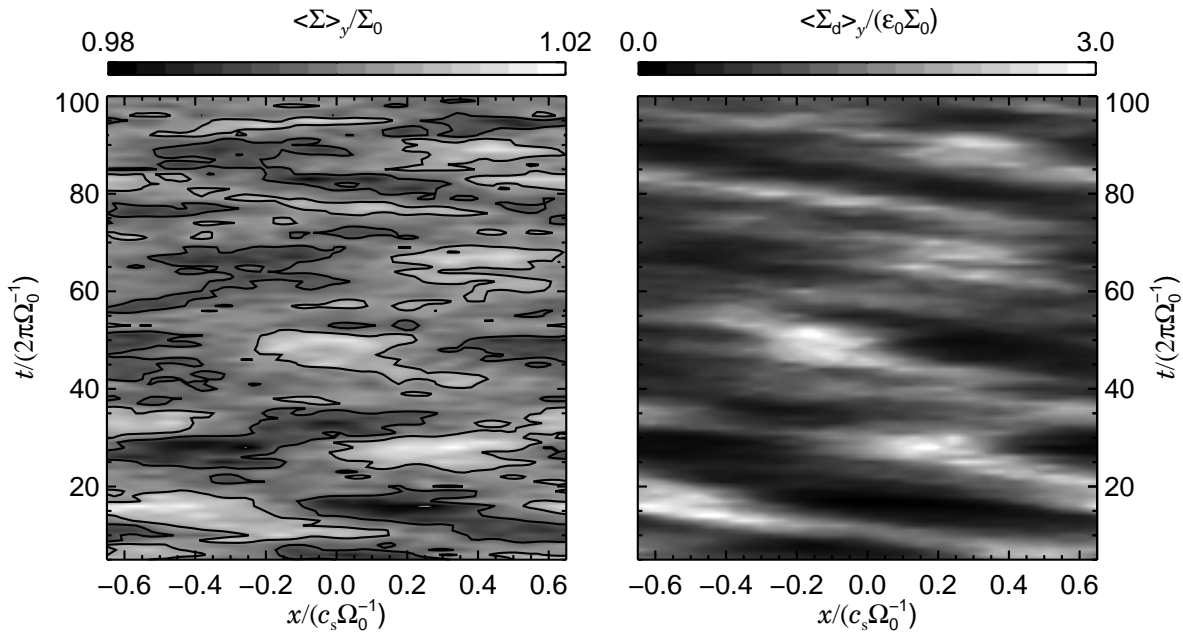,height=20pc}}
\caption{Test particles concentrate in turbulent pressure maxima, from
  3D shearing box simulations of \citet{jkh06}.  The surface
  densities of gas (\emph{left}) and $\taus = 1.0$ solids
  (\emph{right}) are azimuthally-averaged and plotted versus radius
  $x$ (in units of $\hg$) and time $t$ (in orbits).
  Magneto-rotational turbulence drives gas fluctuations, to which
  particles passively respond via drag forces.  The order-unity
  particle density fluctuations overlap the $\sim$$1\%$-scale gas
  density fluctuations---equivalent to the pressure fluctuations for
  the assumed isothermal gas.  Careful inspection shows that solids
  collect slightly downstream (smaller $x$) from local pressure maxima
  since this is where local gradients $\p P/\p x > 0$ cancel the
  global $\p P/\p r < 0$ (see \S\ref{sec_pile}). Reproduced by
  permission of the American Astronomical Society.}
\label{fig:passiveconc}
\end{figure}

It certainly helps to have a disk---or regions of the disk---with 
an abundance of solids relative to gas greater than that given by solar
composition.  The dust-to-gas column density ratio $\Sigmap/\Sigmag$
may need to be supersolar by factors of 2--10 (\S\ref{sec:thin}; see
also \citealt{johansen07}) for planetesimals to form.  The
sharp rise in exoplanet detection rates with host star metallicity
\citep{fv05} reinforces the importance of initial dust abundance.
Disks can also become enriched with time, as pile-ups due to radial
drifts and photoevaporation of gas increase $\Sigmap/\Sigmag$
(Figure \ref{fig:enrich}).  Observations of gas giants are
consistent with their origin in metal-enriched disks.  Bulk metallicities
range from $\sim$$4 \times$ solar (Jupiter; \S\ref{sec_intro}) to
$\sim$$50\times$ solar (HD 141569; \citealt{sato}).

Two popular misconceptions deserve correction.  First, self-gravity
need not be comparable to stellar tides before its effects
are felt. Gas drag on particles, abetted by arbitrarily weak
self-gravity, collects particles into overdense axisymmetric rings
(\S\ref{sec:criteria}). Second, gas turbulence does not necessarily
obstruct GI. There is growing evidence that
magneto-rotational turbulence creates long-lived pressure maxima that
can trap particles (\S\ref{sec_pile}, \S\ref{sec:turb}).

\begin{figure}
\centerline{\psfig{figure=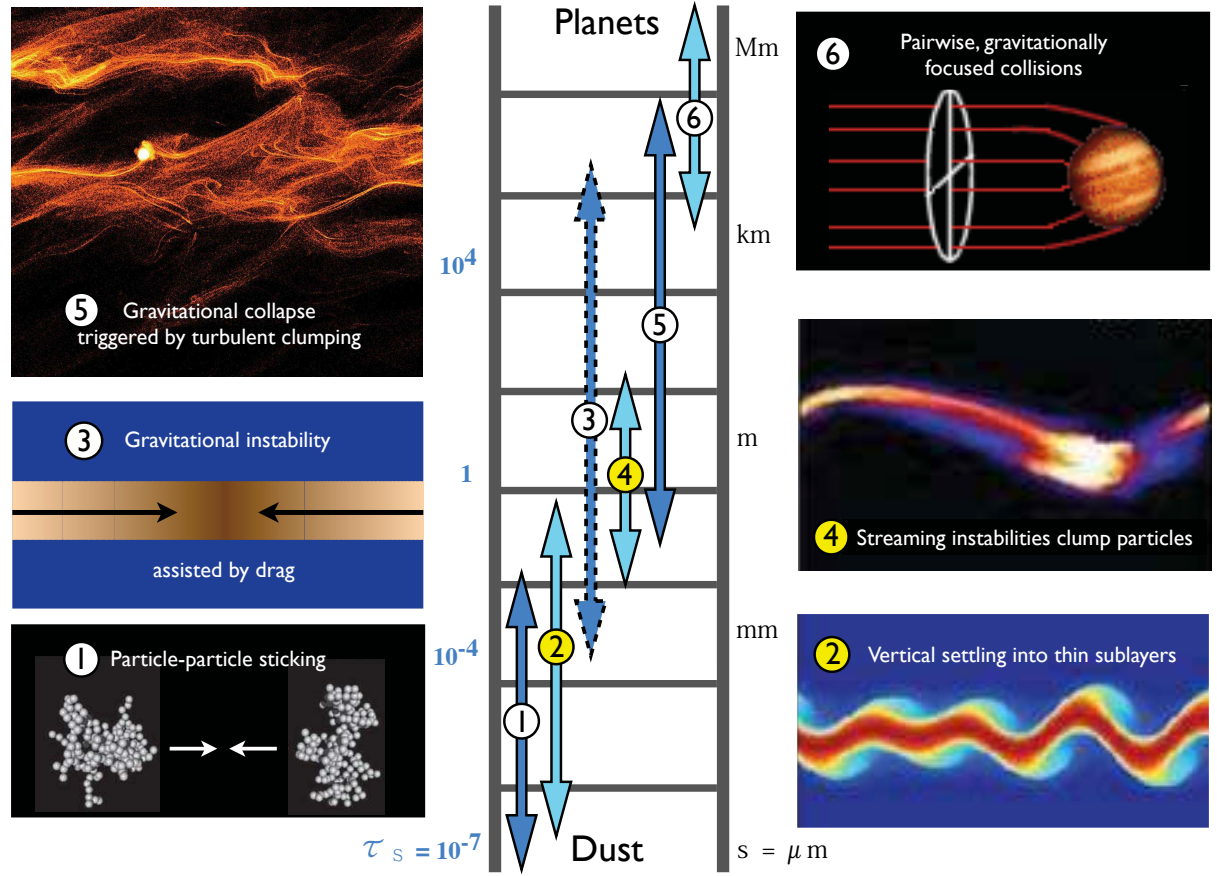,height=30pc}}
\caption{Scaling the size ladder from dust to planets.  Physical
  processes that either grow (labeled with white circles) or
  concentrate solids (yellow) are illustrated. Numbers advance from
  the earliest to latest stages of planet formation.  The relevance of
  a given mechanism tends to be restricted to a certain range of
  particle sizes, indicated crudely by the arrows on the ladder. The
  ranges shown are subject to debate and actively researched.  The
  least well explored is shown by a dashed arrow: drag-assisted GI
  (\S\ref{sec:draggi}). Dimensionless stopping times $\taus \equiv
  \OmegaK \ts$ are shown for $r = 1$ AU; some processes are
  aerodynamic and depend more on $\taus$ than on particle size.}
\label{fig:ladder}
\end{figure}

\begin{figure}
\centerline{\psfig{figure=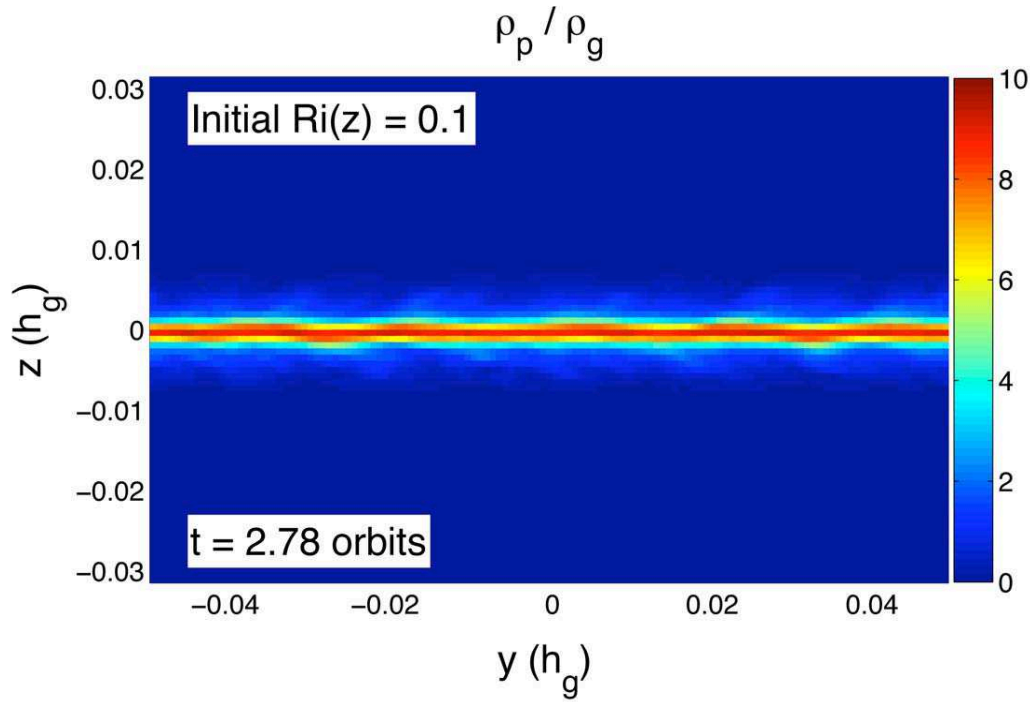,height=23pc}}
\caption{Dusty sublayer undergoing a Kelvin-Helmholtz-like
  instability, taken from a 3D shearing box simulation by Lee et
  al.~(in prep.), using the anelastic code of \citet{b09}. The
  Richardson number is set initially to a constant $\Ri = 0.1$
  \citep{sek98}, with a midplane dust-to-gas ratio of 10.  Larger
  initial values for $\Ri \gtrsim 1$ produce no instability for at
  least 10 orbits.  The vertical shearing parameter $\eta \vk / \cg =
  0.025$.  The size of the shearing box is $(L_x, L_y, L_z) =
  (12.8,6.35,8) z_{\rm max}$, where $z_{\rm max}$ is the maximum
  height of the dust layer \citep[e.g.,][]{c08}, and the number of
  grid points is $(N_x,N_y,N_z) = (128,32,128)$.  }
\label{fig:khi}
\end{figure}

\begin{figure}
\centerline{\psfig{figure=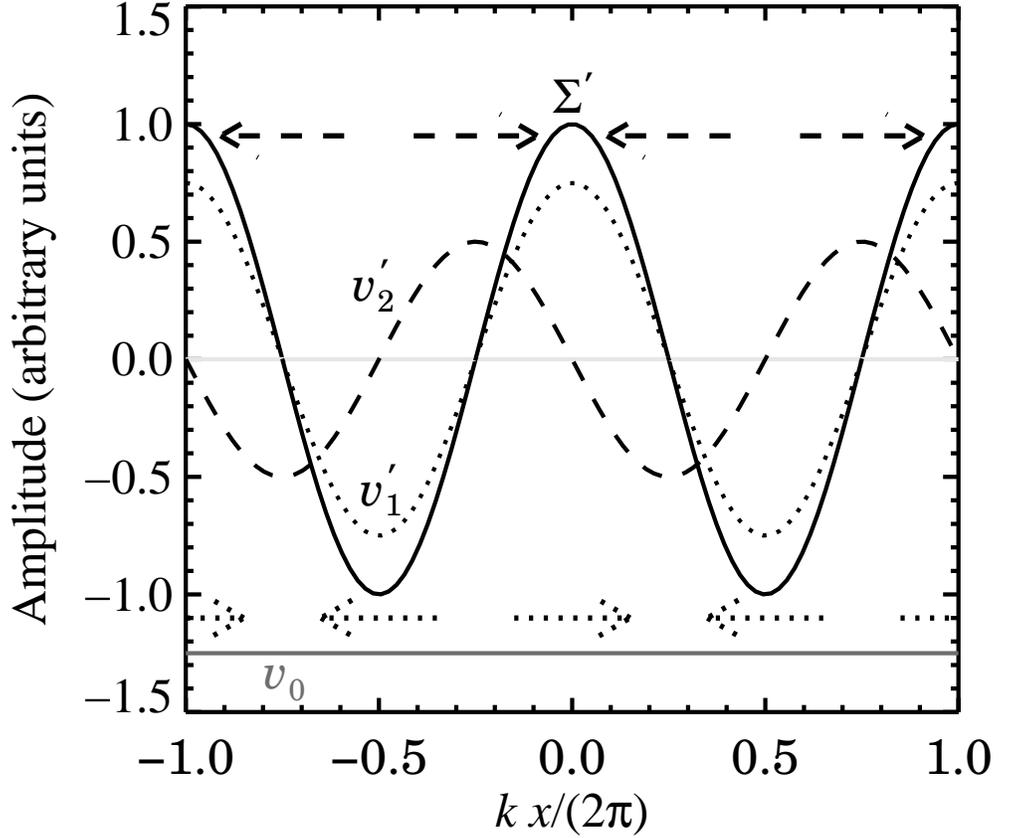,height=30pc}}
\caption{Mechanism behind particle clumping in drag instabilities,
  using the toy model of \citet{gp00}.  The equilibrium inward drift
  speed $v_0$ (grey line) is constant with local radial coordinate
  $x$.  For linear modes, a surface density perturbation $\Sigma'$
  (solid curve) of wavelength $2\pi/k$ has a related velocity
  perturbation $v'$.  To first order in $\delta_\nu \equiv d \ln
  \nu_{\rm d}/d\ln \Sigma > 0$
  (eqs.~\ref{eq:toy2}--\ref{eq:toyexpand}), overdense regions slow
  their drift ($v_1'$, dotted curve).  Since resultant mass fluxes (dotted
  arrows) point toward density zeroes, there is no amplification of
  $\Sigma'$ at this order; the mode merely oscillates. The
  second-order response $v_2'$ (dashed curve) arises from drag forces
  acting on the first-order flow, and is phase-shifted relative to
  $\Sigma'$.  There are now mass fluxes (dashed arrows) that amplify
  the original density perturbation; the mode is overstable.}
\label{fig:DI}
\end{figure}

The above concentration and enrichment mechanisms---combined with vertical
settling (Figure \ref{fig:enrich})---can eventually lead to
$\rhop/\rhog \sim 1$. This condition presents a much lower hurdle 
than the Roche criterion, or even Toomre's $Q$-criterion.
Having as much inertia in particles as in gas paves the way for collective drag
effects---notably the streaming instability---that can clump particles
still further, all the way to Roche density.  Figure \ref{fig:ladder}
illustrates the size ladder that dust must climb to form planets,
identifying key processes.

\begin{figure}
\centerline{\hbox{
\psfig{figure=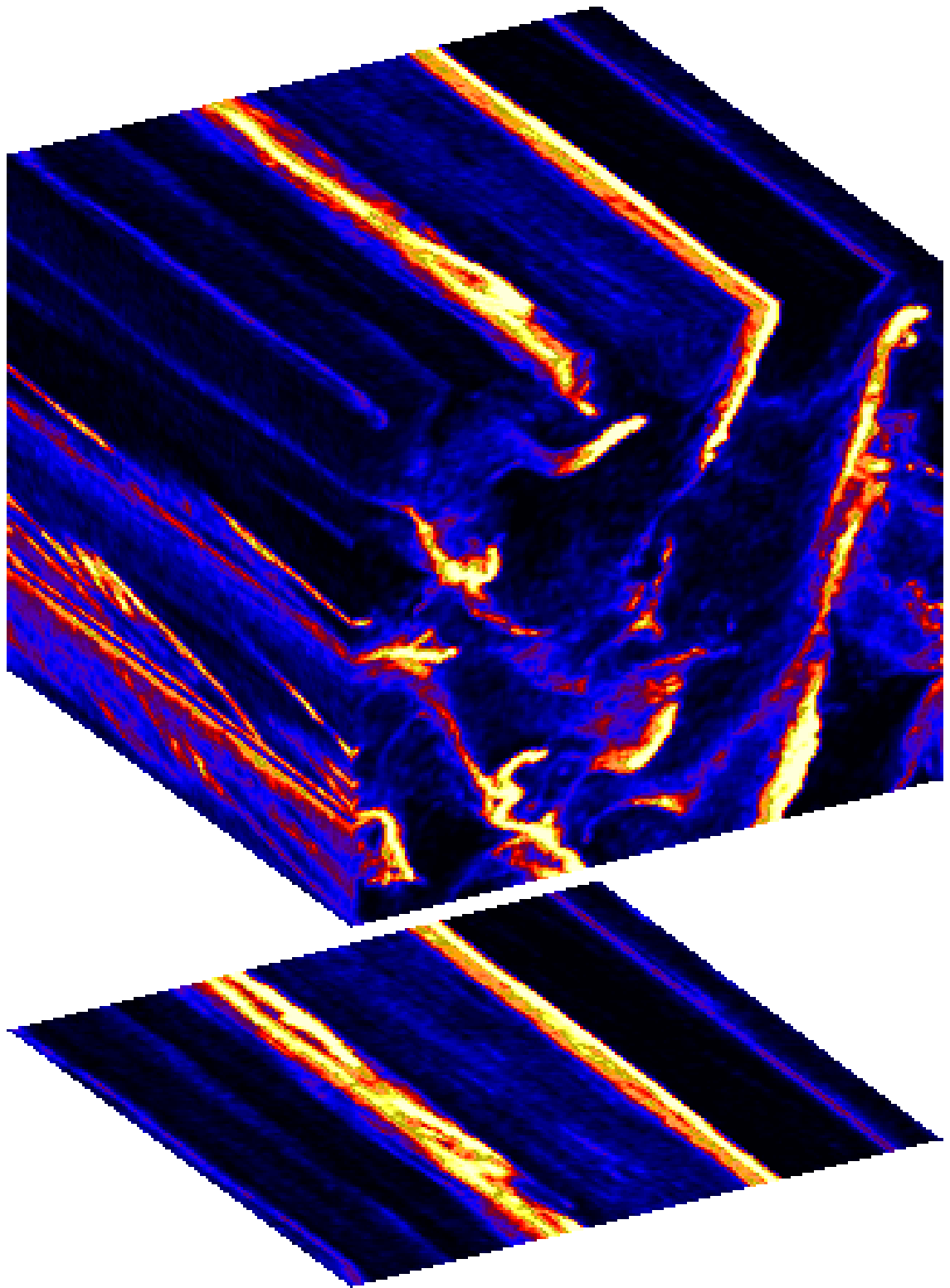,height=20pc}
\psfig{figure=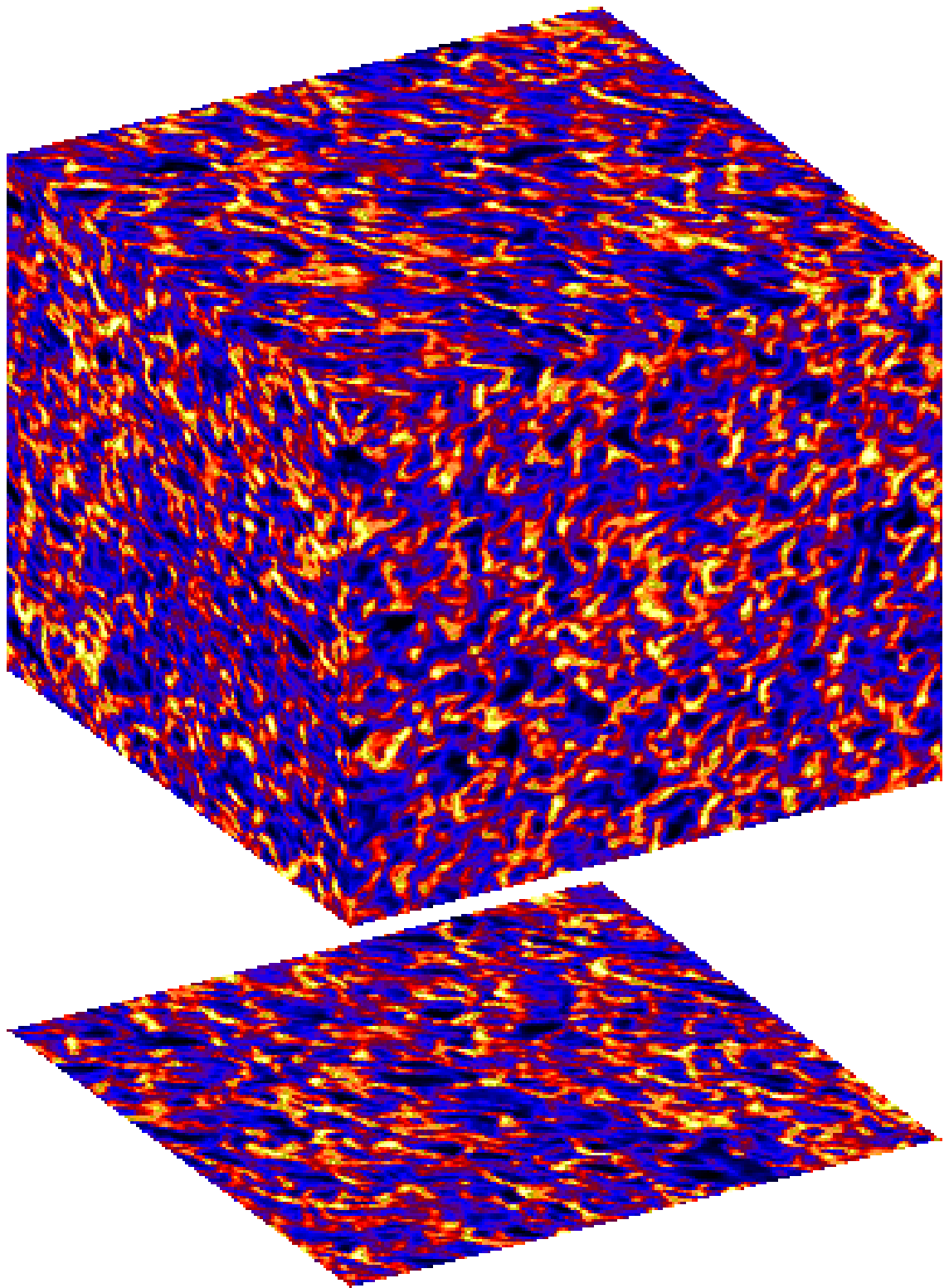,height=20pc}
}}
\caption{Saturated states of the streaming instability, from
  unstratified simulations of mutually coupled particles and gas by
  \citet{jy07}. Shearing box axes are oriented with radius running to
  the right, azimuth to the left, and vertical straight up.  For the
  simulation on the left, $\taus = 1$ and $\langle \rhop \rangle /
  \langle \rhog \rangle = 0.2$; for the right, $\taus = 0.1$ and
  $\langle \rhop \rangle / \langle \rhog \rangle = 1$.  Colors show
  $\rhop$ at the sides of the box, with black corresponding to zero
  density and bright yellow corresponding to densities up to $\sim$$30
  \rhog$ (left) or $\sim$$10\rhog$ (right). Both snapshots are taken
  tens of orbits after the SI has saturated, when turbulence is fully
  developed. For $\taus = 1$, clumps are long-lived, so that Keplerian
  shear smears them into rings. Clumps of more tightly coupled
  particles are shorter-lived. Higher resolution simulations, which
  are feasible in 2D, produce still stronger clumping.  Reproduced by
  permission of the American Astronomical Society.}
\label{fig:SI3D}
\end{figure}

\begin{figure}
\centerline{\psfig{figure=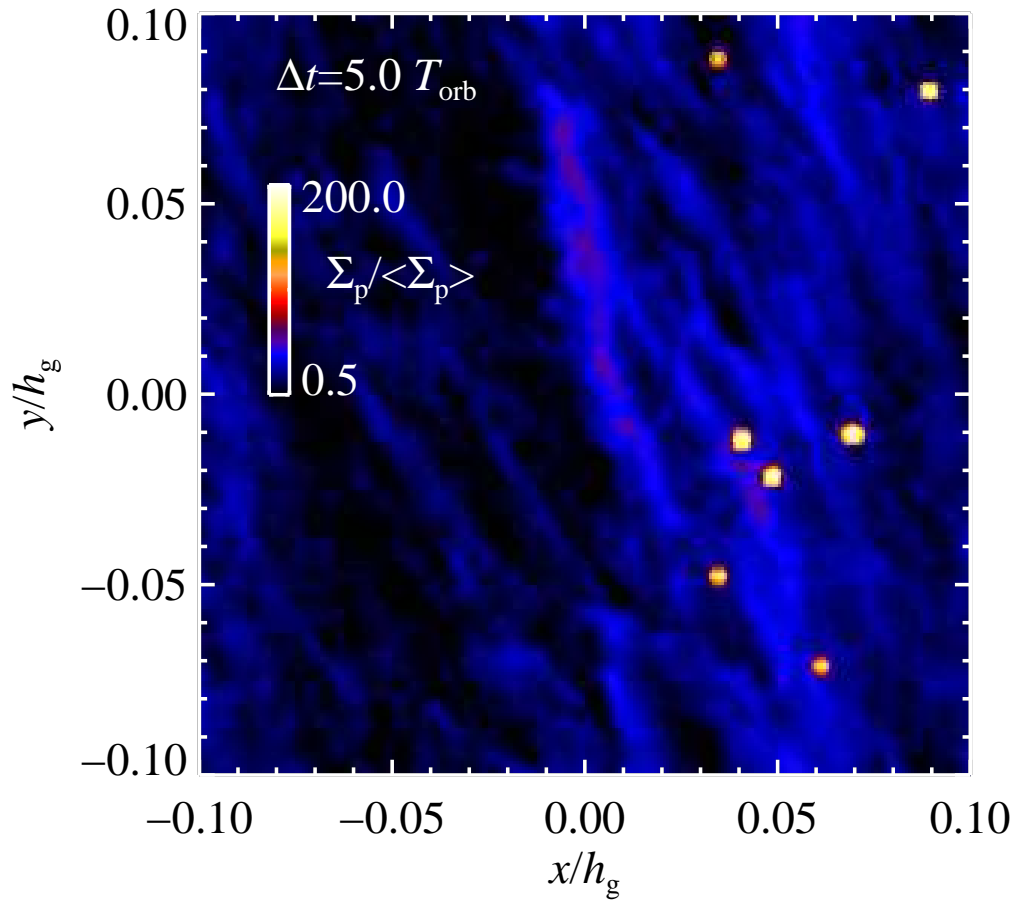,height=30pc}}
\caption{Particle column density $\Sigmap$, showing the formation of
  seven gravitationally bound clumps in a 3D, vertically stratified,
  shearing box simulation of unmagnetized gas and superparticles with
  $\taus = 0.1$--0.4 \citep{jym}. The box-averaged $\langle \Sigmap \rangle/
  \langle \Sigmag \rangle = 0.02$.  The $x (y)$ axis is parallel to
  the radial (azimuthal) direction, and measured in units of the gas
  scale height.  The simulation first evolved for
  40 orbits without self-gravity, during which particles settled to
  the midplane and triggered vertical shearing and streaming
  instabilities; strong clumping resulted.  This snapshot is taken 5
  orbits after self-gravity was turned on. The bound fragments contain
  $\sim$20\% of the total mass in solids; each has a mass comparable
  to a compact planetesimal having a size 100--200 km.}
\label{fig:7sisters}
\end{figure}

Below we chart possible courses for future work:

\begin{enumerate}
\item Chondrules---arguably the building blocks of first-generation
  planetesimals---were flash-heated in an extremely dust-rich
  environment, with $\rhop$ possibly exceeding $\rho^\ast$
  (\S\ref{sec_intro}).  That none of the concentration
  mechanisms discussed in this review (see \S\ref{sec:turbcon})
  explains how they were heated indicates our story may be
  incomplete.  In shock models for chondrules, the isothermal
  post-shock gas density exceeds the pre-shock density by the square
  of the shock Mach number, a factor of about 30---just enough for
  $\rhog \approx \rho^{\ast}$.  This opens the exciting possibility of
  a violent origin for both chondrules and planetesimals, though leaves
  unidentified the source of such strong shocks.

\item To what extent does dust settle to the disk midplane?  The answer 
  depends on stirring by turbulence, whether driven by the solids or 
  externally imposed.  Common idealizations---notably that the dust 
  density profile in otherwise laminar disks has constant Richardson number $\Ri
  \simeq 1/4$---hold only approximately at best, as shown by
  simulations performed in the limit that dust and gas are perfectly
  coupled \citep{c08,b09}. Long-term evolution of the dust distribution 
  requires relaxing the perfect coupling
  assumption, so that dust can sediment out of
  gas. Novel numerical approaches such as implicit time evolution and
  multiple time stepping will need development. The wide separation
  between stopping, orbital, and sedimentation timescales could reveal
  new phenomena.  There may not even be a steady state to which dust
  settles.  In laminar disks, the dust distribution may cycle through
  long periods of settling interrupted by brief bursts of
  nonuniform turbulence wherever $\Ri$ drops below unity.
 
\item Simulations that include the streaming instability
  \citep[e.g.,][]{johansen07} find that particles moderately coupled
  to gas---having stopping times $\taus \geq 0.1$---clump strongly
  enough to undergo gravitational
  collapse. 
  Does the same conclusion apply to yet smaller solids, increasing the
  overlap between process 1---growth by sticking---and process 3 in
  Figure \ref{fig:ladder}? Pessimistically, smaller solids might not
  settle to a layer with $\rhop/\rhog \gtrsim 1$ where streaming
  instabilities are most effective.  Also clump lifetimes decrease
  with decreasing stopping time (Figure \ref{fig:SI3D},
  \S\ref{sec:turbcon}).  Optimistically, particle concentration is
  observed in current simulations to increase with finer grid
  resolution, with no convergence as yet.  An adaptive mesh applied to
  a shearing box would be a powerful tool to track the densest
  particle clumps.  \\

  Together points 2 and 3 emphasize that elucidating the respective
  roles of sticking and self-gravity requires bridging our
  understanding of particle dynamics from the aerodynamically
  well-coupled to the marginally decoupled.
  
\item The post-Roche evolution of particle clusters is largely
  unexplored. A cluster could fragment into multiple ones, much as
  interstellar gas clouds undergo hierarchical fragmentation as the
  Jeans mass decreases with higher density.  The number of fragments
  could be large if the initial overdensity is an azimuthally extended
  ringlet. A record of the initial mass function for planetesimals
  might even be preserved today, in the mass functions of asteroids
  and Kuiper belt objects \citep{morbi09}. Extremely widely separated
  binaries \citep{petit08} may be gravitationally collapsed clusters
  that fissioned.

\end{enumerate}

We thank Xylar Asay-Davis, Joe Barranco, Jeff Cuzzi, Josh Eisner,
Anders Johansen, Aaron Lee, and Paolo Padoan for discussions.
J\"{u}rgen Blum, Anders Johansen, Alessandro Morbidelli, Gordon
Ogilvie, and Jeff Oishi provided helpful and encouraging remarks on
draft versions of this manuscript.  This review owes its origin to the
lunchtime talk series organized through the Berkeley Center for
Integrative Planetary Sciences (CIPS).  This work was supported in
part by National Science Foundation grant AST-0507805.




\bibliographystyle{FluidMechanics}


\begin{thebibliography}{}
\expandafter\ifx\csname natexlab\endcsname\relax\def\natexlab#1{#1}\fi

\bibitem[{{Adachi} et~al.(1976){Adachi}, {Hayashi} \& {Nakazawa}}]{ahn76}
{Adachi} I, {Hayashi} C, {Nakazawa} K. 1976.
\newblock {The gas drag effect on the elliptical motion of a solid body in the
  primordial solar nebula.}
\newblock \textit{Prog.~Theor.~Phys.} 56:1756--1771

\bibitem[{{Alexander} et~al.(2008){Alexander}, {Grossman}, {Ebel} \&
  {Ciesla}}]{alex08}
{Alexander} CMO, {Grossman} JN, {Ebel} DS, {Ciesla} FJ. 2008.
\newblock {The Formation Conditions of Chondrules and Chondrites}.
\newblock \textit{Science} 320:1617--

\bibitem[{{Andrews} \& {Williams}(2005)}]{aw05}
{Andrews} SM, {Williams} JP. 2005.
\newblock {Circumstellar Dust Disks in Taurus-Auriga: The Submillimeter
  Perspective}.
\newblock \textit{Astrophysical Journal} 631:1134--1160

\bibitem[{{Bai} \& {Goodman}(2009)}]{bai}
{Bai} XN, {Goodman} J. 2009.
\newblock {Heat and Dust in Active Layers of Protostellar Disks}.
\newblock \textit{ArXiv e-prints}

\bibitem[{{Balbus}(2009)}]{balbus09}
{Balbus} SA. 2009.
\newblock {Magnetohydrodynamics of Protostellar Disks}.
\newblock \textit{ArXiv e-prints}

\bibitem[{{Balsara} et~al.(2009){Balsara}, {Tilley}, {Rettig} \&
  {Brittain}}]{bal08}
{Balsara} DS, {Tilley} DA, {Rettig} T, {Brittain} SD. 2009.
\newblock {Dust settling in magnetorotationally driven turbulent discs - I.
  Numerical methods and evidence for a vigorous streaming instability}.
\newblock \textit{MNRAS} :804--+

\bibitem[{{Barranco}(2009)}]{b09}
{Barranco} JA. 2009.
\newblock {Three-Dimensional Simulations of Kelvin-Helmholtz Instability in
  Settled Dust Layers in Protoplanetary Disks}.
\newblock \textit{Astrophysical Journal} 691:907--921

\bibitem[{{Bec} et~al.(2007){Bec}, {Biferale}, {Cencini}, {Lanotte},
  {Musacchio} \& {Toschi}}]{bec2007}
{Bec} J, {Biferale} L, {Cencini} M, {Lanotte} A, {Musacchio} S, {Toschi} F.
  2007.
\newblock {Heavy Particle Concentration in Turbulence at Dissipative and
  Inertial Scales}.
\newblock \textit{Physical Review Letters} 98:084502--+

\bibitem[{{Binney} \& {Tremaine}(2008)}]{bt}
{Binney} J, {Tremaine} S. 2008.
\newblock \textit{{Galactic Dynamics: Second Edition}}.
\newblock Princeton University Press

\bibitem[{{Blum} \& {Wurm}(2008)}]{blumwurm08}
{Blum} J, {Wurm} G. 2008.
\newblock {The Growth Mechanisms of Macroscopic Bodies in Protoplanetary
  Disks}.
\newblock \textit{Annual Review of Astronomy and Astrophysics} 46:21--56

\bibitem[{{Burrows} et~al.(1996){Burrows}, {Stapelfeldt}, {Watson}, {Krist},
  {Ballester} et~al.}]{burrows}
{Burrows} CJ, {Stapelfeldt} KR, {Watson} AM, {Krist} JE, {Ballester} GE, et~al.
  1996.
\newblock {Hubble Space Telescope Observations of the Disk and Jet of HH 30}.
\newblock \textit{Astrophysical Journal} 473:437--+

\bibitem[{{Calvet} \& {Gullbring}(1998)}]{calvet98}
{Calvet} N, {Gullbring} E. 1998.
\newblock {The Structure and Emission of the Accretion Shock in T Tauri Stars}.
\newblock \textit{Astrophysical Journal} 509:802--818

\bibitem[{{Calvet} et~al.(2000){Calvet}, {Hartmann} \& {Strom}}]{calvet00}
{Calvet} N, {Hartmann} L, {Strom} SE. 2000.
\newblock {Evolution of Disk Accretion}.
\newblock \textit{Protostars and Planets IV} :377--+

\bibitem[{{Cameron}(1973)}]{cameron}
{Cameron} AGW. 1973.
\newblock {Accumulation processes in the primitive solar nebula}.
\newblock \textit{Icarus} 18:407--450

\bibitem[{{Carballido} et~al.(2006){Carballido}, {Fromang} \&
  {Papaloizou}}]{carb}
{Carballido} A, {Fromang} S, {Papaloizou} J. 2006.
\newblock {Mid-plane sedimentation of large solid bodies in turbulent
  protoplanetary discs}.
\newblock \textit{Monthly Notices of the Royal Astronomical Society}
  373:1633--1640

\bibitem[{{Chandrasekhar}(1987)}]{chandrasek87}
{Chandrasekhar} S. 1987.
\newblock \textit{{Ellipsoidal figures of equilibrium}}.
\newblock Dover

\bibitem[{{Chapman} \& {Cowling}(1970)}]{chapcow}
{Chapman} S, {Cowling} TG. 1970.
\newblock \textit{{The mathematical theory of non-uniform gases. an account of
  the kinetic theory of viscosity, thermal conduction and diffusion in gases}}.
\newblock Cambridge University Press

\bibitem[{{Chavanis}(2000)}]{chav00}
{Chavanis} PH. 2000.
\newblock {Trapping of dust by coherent vortices in the solar nebula}.
\newblock \textit{Astronomy and Astrophysics} 356:1089--1111

\bibitem[{{Chiang}(2008)}]{c08}
{Chiang} E. 2008.
\newblock {Vertical Shearing Instabilities in Radially Shearing Disks: The
  Dustiest Layers of the Protoplanetary Nebula}.
\newblock \textit{Astrophysical Journal} 675:1549--1558

\bibitem[{{Chiang} \& {Murray-Clay}(2007)}]{cmc}
{Chiang} E, {Murray-Clay} R. 2007.
\newblock {Inside-out evacuation of transitional protoplanetary discs by the
  magneto-rotational instability}.
\newblock \textit{Nature Physics} 3:604--608

\bibitem[{{Chiang} \& {Goldreich}(1997)}]{cg97}
{Chiang} EI, {Goldreich} P. 1997.
\newblock {Spectral Energy Distributions of T Tauri Stars with Passive
  Circumstellar Disks}.
\newblock \textit{Astrophysical Journal} 490:368--+

\bibitem[{{Chiang} et~al.(2001){Chiang}, {Joung}, {Creech-Eakman}, {Qi},
  {Kessler} et~al.}]{c01}
{Chiang} EI, {Joung} MK, {Creech-Eakman} MJ, {Qi} C, {Kessler} JE, et~al. 2001.
\newblock {Spectral Energy Distributions of Passive T Tauri and Herbig Ae
  Disks: Grain Mineralogy, Parameter Dependences, and Comparison with Infrared
  Space Observatory LWS Observations}.
\newblock \textit{Astrophysical Journal} 547:1077--1089

\bibitem[{{Chokshi} et~al.(1993){Chokshi}, {Tielens} \& {Hollenbach}}]{chokshi}
{Chokshi} A, {Tielens} AGGM, {Hollenbach} D. 1993.
\newblock {Dust coagulation}.
\newblock \textit{Astrophysical Journal} 407:806--819

\bibitem[{{Ciesla} \& {Cuzzi}(2006)}]{cc06}
{Ciesla} FJ, {Cuzzi} JN. 2006.
\newblock {The evolution of the water distribution in a viscous protoplanetary
  disk}.
\newblock \textit{Icarus} 181:178--204

\bibitem[{{Coradini} et~al.(1981){Coradini}, {Magni} \& {Federico}}]{coradini}
{Coradini} A, {Magni} G, {Federico} C. 1981.
\newblock {Formation of planetesimals in an evolving protoplanetary disk}.
\newblock \textit{Astronomy and Astrophysics} 98:173--185

\bibitem[{{Cuzzi} et~al.(1993){Cuzzi}, {Dobrovolskis} \& {Champney}}]{cuzzi93}
{Cuzzi} JN, {Dobrovolskis} AR, {Champney} JM. 1993.
\newblock {Particle-gas dynamics in the midplane of a protoplanetary nebula}.
\newblock \textit{Icarus} 106:102--+

\bibitem[{{Cuzzi} et~al.(2001){Cuzzi}, {Hogan}, {Paque} \&
  {Dobrovolskis}}]{cuz01}
{Cuzzi} JN, {Hogan} RC, {Paque} JM, {Dobrovolskis} AR. 2001.
\newblock {Size-selective Concentration of Chondrules and Other Small Particles
  in Protoplanetary Nebula Turbulence}.
\newblock \textit{Astrophysical Journal} 546:496--508

\bibitem[{{Cuzzi} et~al.(2008){Cuzzi}, {Hogan} \& {Shariff}}]{chs08}
{Cuzzi} JN, {Hogan} RC, {Shariff} K. 2008.
\newblock {Toward Planetesimals: Dense Chondrule Clumps in the Protoplanetary
  Nebula}.
\newblock \textit{Astrophysical Journal} 687:1432--1447

\bibitem[{{D'Alessio} et~al.(2001){D'Alessio}, {Calvet} \& {Hartmann}}]{daless}
{D'Alessio} P, {Calvet} N, {Hartmann} L. 2001.
\newblock {Accretion Disks around Young Objects. III. Grain Growth}.
\newblock \textit{Astrophysical Journal} 553:321--334

\bibitem[{{Davis} et~al.(2009){Davis}, {Stone} \& {Pessah}}]{davis}
{Davis} SW, {Stone} JM, {Pessah} ME. 2009.
\newblock {Sustained Magnetorotational Turbulence in Local Simulations of
  Stratified Disks with Zero Net Magnetic Flux}.
\newblock \textit{ArXiv e-prints}

\bibitem[{{Desch} et~al.(2005){Desch}, {Ciesla}, {Hood} \&
  {Nakamoto}}]{deschetal}
{Desch} SJ, {Ciesla} FJ, {Hood} LL, {Nakamoto} T. 2005.
\newblock {Heating of Chondritic Materials in Solar Nebula Shocks}.
\newblock In \textit{Chondrites and the Protoplanetary Disk}, eds. AN~{Krot},
  ERD {Scott}, B~{Reipurth}, vol. 341 of \textit{Astronomical Society of the
  Pacific Conference Series}

\bibitem[{{Dominik} \& {Tielens}(1997)}]{dt97}
{Dominik} C, {Tielens} AGGM. 1997.
\newblock {The Physics of Dust Coagulation and the Structure of Dust Aggregates
  in Space}.
\newblock \textit{Astrophysical Journal} 480:647--+

\bibitem[{{Drazin} \& {Reid}(2004)}]{drazin}
{Drazin} PG, {Reid} WH. 2004.
\newblock \textit{{Hydrodynamic Stability}}.
\newblock Cambridge University Press

\bibitem[{{Dubrulle} et~al.(1995){Dubrulle}, {Morfill} \& {Sterzik}}]{dms95}
{Dubrulle} B, {Morfill} G, {Sterzik} M. 1995.
\newblock {The dust subdisk in the protoplanetary nebula}.
\newblock \textit{Icarus} 114:237--246

\bibitem[{{Dullemond} \& {Dominik}(2005)}]{dd05}
{Dullemond} CP, {Dominik} C. 2005.
\newblock {Dust coagulation in protoplanetary disks: A rapid depletion of small
  grains}.
\newblock \textit{Astronomy and Astrophysics} 434:971--986

\bibitem[{{Dullemond} et~al.(2007){Dullemond}, {Hollenbach}, {Kamp} \&
  {D'Alessio}}]{dulle}
{Dullemond} CP, {Hollenbach} D, {Kamp} I, {D'Alessio} P. 2007.
\newblock {Models of the Structure and Evolution of Protoplanetary Disks}.
\newblock In \textit{Protostars and Planets V}, eds. B~{Reipurth}, D~{Jewitt},
  K~{Keil}

\bibitem[{{Eaton} \& {Fessler}(1994)}]{eatonfessler}
{Eaton} JK, {Fessler} JR. 1994.
\newblock {Preferential Concentration of Particles by Turbulence}.
\newblock \textit{International Journal of Multiphase Flow} 20:169--209

\bibitem[{{Epstein}(1924)}]{epstein}
{Epstein} PS. 1924.
\newblock {On the Resistance Experienced by Spheres in their Motion through
  Gases}.
\newblock \textit{Physical Review} 23:710--733

\bibitem[{{Ercolano} et~al.(2009){Ercolano}, {Clarke} \& {Drake}}]{ercolano}
{Ercolano} B, {Clarke} CJ, {Drake} JJ. 2009.
\newblock {X-Ray Irradiated Protoplanetary Disk Atmospheres. II. Predictions
  from Models in Hydrostatic Equilibrium}.
\newblock \textit{Astrophysical Journal} 699:1639--1649

\bibitem[{{Fischer} \& {Valenti}(2005)}]{fv05}
{Fischer} DA, {Valenti} J. 2005.
\newblock {The Planet-Metallicity Correlation}.
\newblock \textit{Astrophysical Journal} 622:1102--1117

\bibitem[{{Fleming} \& {Stone}(2003)}]{fs03}
{Fleming} T, {Stone} JM. 2003.
\newblock {Local Magnetohydrodynamic Models of Layered Accretion Disks}.
\newblock \textit{Astrophysical Journal} 585:908--920

\bibitem[{{Frank} et~al.(2002){Frank}, {King} \& {Raine}}]{fkr}
{Frank} J, {King} A, {Raine} DJ. 2002.
\newblock \textit{{Accretion Power in Astrophysics: Third Edition}}.
\newblock Cambridge University Press

\bibitem[{{Frisch}(1996)}]{frisch}
{Frisch} U. 1996.
\newblock \textit{{Turbulence}}.
\newblock Cambridge University Press

\bibitem[{{Fromang} \& {Nelson}(2005)}]{fn05}
{Fromang} S, {Nelson} RP. 2005.
\newblock {On the accumulation of solid bodies in global turbulent
  protoplanetary disc models}.
\newblock \textit{MNRAS} 364:L81--L85

\bibitem[{{Fromang} et~al.(2009){Fromang}, {Papaloizou}, {Lesur} \&
  {Heinemann}}]{fromang09}
{Fromang} S, {Papaloizou} J, {Lesur} G, {Heinemann} T. 2009.
\newblock {Numerical Simulations of MHD Turbulence in Accretion Disks}.
\newblock In \textit{Astronomical Society of the Pacific Conference Series},
  eds. NV~{Pogorelov}, E~{Audit}, P~{Colella}, GP~{Zank}, vol. 406 of
  \textit{Astronomical Society of the Pacific Conference Series}

\bibitem[{{Gammie}(1996)}]{gammie96}
{Gammie} CF. 1996.
\newblock {Layered Accretion in T Tauri Disks}.
\newblock \textit{Astrophysical Journal} 457:355--+

\bibitem[{{Gammie}(2001)}]{gam01}
{Gammie} CF. 2001.
\newblock {Nonlinear Outcome of Gravitational Instability in Cooling, Gaseous
  Disks}.
\newblock \textit{Astrophysical Journal} 553:174--183

\bibitem[{{Garaud} \& {Lin}(2004)}]{garaud}
{Garaud} P, {Lin} DNC. 2004.
\newblock {On the Evolution and Stability of a Protoplanetary Disk Dust Layer}.
\newblock \textit{Astrophysical Journal} 608:1050--1075

\bibitem[{{Goldreich} et~al.(2004){Goldreich}, {Lithwick} \& {Sari}}]{g04}
{Goldreich} P, {Lithwick} Y, {Sari} R. 2004.
\newblock {Planet Formation by Coagulation: A Focus on Uranus and Neptune}.
\newblock \textit{Annual Review of Astronomy and Astrophysics} 42:549--601

\bibitem[{{Goldreich} \& {Lynden-Bell}(1965)}]{glb}
{Goldreich} P, {Lynden-Bell} D. 1965.
\newblock {II. Spiral arms as sheared gravitational instabilities}.
\newblock \textit{Monthly Notices of the Royal Astronomical Society} 130:125--+

\bibitem[{{Goldreich} \& {Ward}(1973)}]{gw}
{Goldreich} P, {Ward} WR. 1973.
\newblock {The Formation of Planetesimals}.
\newblock \textit{Astrophysical Journal} 183:1051--1062

\bibitem[{{G{\'o}mez} \& {Ostriker}(2005)}]{go}
{G{\'o}mez} GC, {Ostriker} EC. 2005.
\newblock {The Effect of the Coriolis Force on Kelvin-Helmholtz-driven Mixing
  in Protoplanetary Disks}.
\newblock \textit{Astrophysical Journal} 630:1093--1106

\bibitem[{{Gooding} \& {Keil}(1981)}]{goodingkeil}
{Gooding} JL, {Keil} K. 1981.
\newblock {Relative abundances of chondrule primary textural types in ordinary
  chondrites and their bearing on conditions of chondrule formation}.
\newblock \textit{Meteoritics} 16:17--43

\bibitem[{{Goodman} \& {Pindor}(2000)}]{gp00}
{Goodman} J, {Pindor} B. 2000.
\newblock {Secular Instability and Planetesimal Formation in the Dust Layer}.
\newblock \textit{Icarus} 148:537--549

\bibitem[{{Gorti} \& {Hollenbach}(2009)}]{gorti}
{Gorti} U, {Hollenbach} D. 2009.
\newblock {Photoevaporation of Circumstellar Disks By Far-Ultraviolet,
  Extreme-Ultraviolet and X-Ray Radiation from the Central Star}.
\newblock \textit{Astrophysical Journal} 690:1539--1552

\bibitem[{{Greenberg} et~al.(1991){Greenberg}, {Bottke}, {Carusi} \&
  {Valsecchi}}]{greenberg}
{Greenberg} R, {Bottke} WF, {Carusi} A, {Valsecchi} GB. 1991.
\newblock {Planetary accretion rates - Analytical derivation}.
\newblock \textit{Icarus} 94:98--111

\bibitem[{{Guillot}(2005)}]{guillot}
{Guillot} T. 2005.
\newblock {The Interiors of Giant Planets: Models and Outstanding Questions}.
\newblock \textit{Annual Review of Earth and Planetary Sciences} 33:493--530

\bibitem[{{Guilloteau} et~al.(2008){Guilloteau}, {Dutrey}, {Pety} \&
  {Gueth}}]{guilloteau}
{Guilloteau} S, {Dutrey} A, {Pety} J, {Gueth} F. 2008.
\newblock {Resolving the circumbinary dust disk surrounding HH 30}.
\newblock \textit{Astronomy and Astrophysics} 478:L31--L34

\bibitem[{{Hartmann} et~al.(2006){Hartmann}, {D'Alessio}, {Calvet} \&
  {Muzerolle}}]{hartmann}
{Hartmann} L, {D'Alessio} P, {Calvet} N, {Muzerolle} J. 2006.
\newblock {Why Do T Tauri Disks Accrete?}
\newblock \textit{Astrophysical Journal} 648:484--490

\bibitem[{{Hern{\'a}ndez} et~al.(2008){Hern{\'a}ndez}, {Hartmann}, {Calvet},
  {Jeffries}, {Gutermuth} et~al.}]{hern}
{Hern{\'a}ndez} J, {Hartmann} L, {Calvet} N, {Jeffries} RD, {Gutermuth} R,
  et~al. 2008.
\newblock {A Spitzer View of Protoplanetary Disks in the {$\gamma$} Velorum
  Cluster}.
\newblock \textit{Astrophysical Journal} 686:1195--1208

\bibitem[{{Hewins}(1996)}]{hewins}
{Hewins} RH. 1996.
\newblock {Chondrules and the protoplanetary disk: an overview.}
\newblock In \textit{Chondrules and the Protoplanetary Disk}, eds. RH~{Hewins},
  R~{Jones}, E~{Scott}. Cambridge University Press

\bibitem[{{Hillenbrand}(2005)}]{hill05}
{Hillenbrand} LA. 2005.
\newblock {Observational Constraints on Dust Disk Lifetimes: Implications for
  Planet Formation}.
\newblock \textit{ArXiv Astrophysics e-prints}

\bibitem[{{Hogan} \& {Cuzzi}(2007)}]{hc07}
{Hogan} RC, {Cuzzi} JN. 2007.
\newblock {Cascade model for particle concentration and enstrophy in fully
  developed turbulence with mass-loading feedback}.
\newblock \textit{Physical Review E} 75:056305--+

\bibitem[{{Ida} \& {Lin}(2008)}]{idalin08}
{Ida} S, {Lin} DNC. 2008.
\newblock {Toward a Deterministic Model of Planetary Formation. V. Accumulation
  Near the Ice Line and Super-Earths}.
\newblock \textit{Astrophysical Journal} 685:584--595

\bibitem[{{Ishitsu} et~al.(2009){Ishitsu}, {Inutsuka} \& {Sekiya}}]{ish09}
{Ishitsu} N, {Inutsuka} Si, {Sekiya} M. 2009.
\newblock {Two-fluid Instability of Dust and Gas in the Dust Layer of a
  Protoplanetary Disk}.
\newblock \textit{ArXiv e-prints}

\bibitem[{{Ishitsu} \& {Sekiya}(2003)}]{is}
{Ishitsu} N, {Sekiya} M. 2003.
\newblock {The effects of the tidal force on shear instabilities in the dust
  layer of the solar nebula}.
\newblock \textit{Icarus} 165:181--194

\bibitem[{{Johansen} et~al.(2006{\natexlab{a}}){Johansen}, {Henning} \&
  {Klahr}}]{jhk06}
{Johansen} A, {Henning} T, {Klahr} H. 2006{\natexlab{a}}.
\newblock {Dust Sedimentation and Self-sustained Kelvin-Helmholtz Turbulence in
  Protoplanetary Disk Midplanes}.
\newblock \textit{Astrophysical Journal} 643:1219--1232

\bibitem[{{Johansen} et~al.(2006{\natexlab{b}}){Johansen}, {Klahr} \&
  {Henning}}]{jkh06}
{Johansen} A, {Klahr} H, {Henning} T. 2006{\natexlab{b}}.
\newblock {Gravoturbulent Formation of Planetesimals}.
\newblock \textit{Astrophysical Journal} 636:1121--1134

\bibitem[{{Johansen} et~al.(2007){Johansen}, {Oishi}, {MacLow}, {Klahr},
  {Henning} \& {Youdin}}]{johansen07}
{Johansen} A, {Oishi} JS, {MacLow} MM, {Klahr} H, {Henning} T, {Youdin} A.
  2007.
\newblock {Rapid planetesimal formation in turbulent circumstellar disks}.
\newblock \textit{Nature} 448:1022--1025

\bibitem[{{Johansen} \& {Youdin}(2007)}]{jy07}
{Johansen} A, {Youdin} A. 2007.
\newblock {Protoplanetary Disk Turbulence Driven by the Streaming Instability:
  Nonlinear Saturation and Particle Concentration}.
\newblock \textit{Astrophysical Journal} 662:627--641

\bibitem[{{Johansen} et~al.(2009{\natexlab{a}}){Johansen}, {Youdin} \&
  {Klahr}}]{jyk09}
{Johansen} A, {Youdin} A, {Klahr} H. 2009{\natexlab{a}}.
\newblock {Zonal Flows and Long-lived Axisymmetric Pressure Bumps in
  Magnetorotational Turbulence}.
\newblock \textit{Astrophysical Journal} 697:1269--1289

\bibitem[{{Johansen} et~al.(2009{\natexlab{b}}){Johansen}, {Youdin} \&
  {MacLow}}]{jym}
{Johansen} A, {Youdin} A, {MacLow} MM. 2009{\natexlab{b}}.
\newblock {Particle Clumping in Protoplanetary Disks Depends Strongly on
  Metallicity}.
\newblock \textit{ArXiv e-prints}

\bibitem[{{Jones} et~al.(2008){Jones}, {Butler}, {Wright}, {Marcy}, {Fischer}
  et~al.}]{jones}
{Jones} HRA, {Butler} RP, {Wright} JT, {Marcy} GW, {Fischer} DA, et~al. 2008.
\newblock {A Catalogue of Nearby Exoplanets}.
\newblock In \textit{Precision Spectroscopy in Astrophysics}, eds. NC~{Santos},
  L~{Pasquini}, ACM {Correia}, M~{Romaniello}

\bibitem[{{Kato} et~al.(2009){Kato}, {Nakamura}, {Tandokoro}, {Fujimoto} \&
  {Ida}}]{kato}
{Kato} MT, {Nakamura} K, {Tandokoro} R, {Fujimoto} M, {Ida} S. 2009.
\newblock {Modification of Angular Velocity by Inhomogeneous Magnetorotational
  Instability Growth in Protoplanetary Disks}.
\newblock \textit{Astrophysical Journal} 691:1697--1706

\bibitem[{{Kippenhahn} \& {Weigert}(1990)}]{kip}
{Kippenhahn} R, {Weigert} A. 1990.
\newblock \textit{{Stellar Structure and Evolution}}.
\newblock Springer-Verlag

\bibitem[{{Knobloch} \& {Spruit}(1985)}]{knob}
{Knobloch} E, {Spruit} HC. 1985.
\newblock {Baroclinic instability in the presence of a strong horizontal
  shear}.
\newblock \textit{Geophysical and Astrophysical Fluid Dynamics} 32:197--216

\bibitem[{{Kusaka} et~al.(1970){Kusaka}, {Nakano} \& {Hayashi}}]{kusaka}
{Kusaka} T, {Nakano} T, {Hayashi} C. 1970.
\newblock {Growth of Solid Particles in the Primordial Solar Nebula}.
\newblock \textit{Progress of Theoretical Physics} 44:1580--1595

\bibitem[{{Lesur} \& {Longaretti}(2007)}]{lesur}
{Lesur} G, {Longaretti} PY. 2007.
\newblock {Impact of dimensionless numbers on the efficiency of
  magnetorotational instability induced turbulent transport}.
\newblock \textit{Monthly Notices of the Royal Astronomical Society}
  378:1471--1480

\bibitem[{{Lissauer} et~al.(2009){Lissauer}, {Hubickyj}, {D'Angelo} \&
  {Bodenheimer}}]{liss09}
{Lissauer} JJ, {Hubickyj} O, {D'Angelo} G, {Bodenheimer} P. 2009.
\newblock {Models of Jupiter's growth incorporating thermal and hydrodynamic
  constraints}.
\newblock \textit{Icarus} 199:338--350

\bibitem[{{Lissauer} \& {Stevenson}(2007)}]{liss07}
{Lissauer} JJ, {Stevenson} DJ. 2007.
\newblock {Formation of Giant Planets}.
\newblock In \textit{Protostars and Planets V}, eds. B~{Reipurth}, D~{Jewitt},
  K~{Keil}

\bibitem[{{Lithwick}(2009)}]{lithwick}
{Lithwick} Y. 2009.
\newblock {Formation, Survival, and Destruction of Vortices in Accretion
  Disks}.
\newblock \textit{Astrophysical Journal} 693:85--96

\bibitem[{{Lodders}(2003)}]{lodders}
{Lodders} K. 2003.
\newblock {Solar System Abundances and Condensation Temperatures of the
  Elements}.
\newblock \textit{Astrophysical Journal} 591:1220--1247

\bibitem[{{Malhotra}(1993)}]{malhotra}
{Malhotra} R. 1993.
\newblock {The origin of Pluto's peculiar orbit}.
\newblock \textit{Nature} 365:819--821

\bibitem[{{Maxey}(1987)}]{max87}
{Maxey} MR. 1987.
\newblock {The gravitational settling of aerosol particles in homogeneous
  turbulence and random flow fields}.
\newblock \textit{Journal of Fluid Mechanics} 174:441--465

\bibitem[{{McCabe} et~al.(2003){McCabe}, {Duch{\^e}ne} \& {Ghez}}]{mccabe}
{McCabe} C, {Duch{\^e}ne} G, {Ghez} AM. 2003.
\newblock {The First Detection of Spatially Resolved Mid-Infrared Scattered
  Light from a Protoplanetary Disk}.
\newblock \textit{Astrophysical Journal Letters} 588:L113--L116

\bibitem[{{Michikoshi} et~al.(2007){Michikoshi}, {Inutsuka}, {Kokubo} \&
  {Furuya}}]{michi}
{Michikoshi} S, {Inutsuka} Si, {Kokubo} E, {Furuya} I. 2007.
\newblock {N-Body Simulation of Planetesimal Formation through Gravitational
  Instability of a Dust Layer}.
\newblock \textit{Astrophysical Journal} 657:521--532

\bibitem[{{Morbidelli} et~al.(2009){Morbidelli}, {Bottke}, {Nesvorny} \&
  {Levison}}]{morbi09}
{Morbidelli} A, {Bottke} W, {Nesvorny} D, {Levison} HF. 2009.
\newblock {Asteroids Were Born Big}.
\newblock \textit{ArXiv e-prints}

\bibitem[{{Nakagawa} et~al.(1986){Nakagawa}, {Sekiya} \& {Hayashi}}]{nakagawa}
{Nakagawa} Y, {Sekiya} M, {Hayashi} C. 1986.
\newblock {Settling and growth of dust particles in a laminar phase of a
  low-mass solar nebula}.
\newblock \textit{Icarus} 67:375--390

\bibitem[{{Natta} et~al.(2007){Natta}, {Testi}, {Calvet}, {Henning}, {Waters}
  \& {Wilner}}]{natta}
{Natta} A, {Testi} L, {Calvet} N, {Henning} T, {Waters} R, {Wilner} D. 2007.
\newblock {Dust in Protoplanetary Disks: Properties and Evolution}.
\newblock In \textit{Protostars and Planets V}, eds. B~{Reipurth}, D~{Jewitt},
  K~{Keil}

\bibitem[{{Noh} et~al.(1991){Noh}, {Vishniac} \& {Cochran}}]{noh}
{Noh} H, {Vishniac} ET, {Cochran} WD. 1991.
\newblock {Gravitational instabilities in a proto-planetary disk}.
\newblock \textit{Astrophysical Journal} 383:372--379

\bibitem[{{Oishi} et~al.(2007){Oishi}, {Mac Low} \& {Menou}}]{oishi07}
{Oishi} JS, {Mac Low} MM, {Menou} K. 2007.
\newblock {Turbulent Torques on Protoplanets in a Dead Zone}.
\newblock \textit{Astrophysical Journal} 670:805--819

\bibitem[{{Ormel} \& {Cuzzi}(2007)}]{ormelcuzzi}
{Ormel} CW, {Cuzzi} JN. 2007.
\newblock {Closed-form expressions for particle relative velocities induced by
  turbulence}.
\newblock \textit{Astronomy and Astrophysics} 466:413--420

\bibitem[{{Ormel} et~al.(2007){Ormel}, {Spaans} \& {Tielens}}]{ormel}
{Ormel} CW, {Spaans} M, {Tielens} AGGM. 2007.
\newblock {Dust coagulation in protoplanetary disks: porosity matters}.
\newblock \textit{Astronomy and Astrophysics} 461:215--232

\bibitem[{{Pan} \& {Sari}(2005)}]{pan}
{Pan} M, {Sari} R. 2005.
\newblock {Shaping the Kuiper belt size distribution by shattering large but
  strengthless bodies}.
\newblock \textit{Icarus} 173:342--348

\bibitem[{{Petit} et~al.(2008){Petit}, {Kavelaars}, {Gladman}, {Margot},
  {Nicholson} et~al.}]{petit08}
{Petit} JM, {Kavelaars} JJ, {Gladman} BJ, {Margot} JL, {Nicholson} PD, et~al.
  2008.
\newblock {The Extreme Kuiper Belt Binary 2001 QW$_{322}$}.
\newblock \textit{Science} 322:432--

\bibitem[{{Rice} et~al.(2004){Rice}, {Lodato}, {Pringle}, {Armitage} \&
  {Bonnell}}]{rice}
{Rice} WKM, {Lodato} G, {Pringle} JE, {Armitage} PJ, {Bonnell} IA. 2004.
\newblock {Accelerated planetesimal growth in self-gravitating protoplanetary
  discs}.
\newblock \textit{Monthly Notices of the Royal Astronomical Society}
  355:543--552

\bibitem[{{Safronov}(1969)}]{saf}
{Safronov} VS. 1969.
\newblock \textit{{Evoliutsiia doplanetnogo oblaka.}}
\newblock Nauka Press

\bibitem[{{Sato} et~al.(2005){Sato}, {Fischer}, {Henry}, {Laughlin}, {Butler}
  et~al.}]{sato}
{Sato} B, {Fischer} DA, {Henry} GW, {Laughlin} G, {Butler} RP, et~al. 2005.
\newblock {The N2K Consortium. II. A Transiting Hot Saturn around HD 149026
  with a Large Dense Core}.
\newblock \textit{Astrophysical Journal} 633:465--473

\bibitem[{{Sekiya}(1983)}]{sek83}
{Sekiya} M. 1983.
\newblock {Gravitational instabilities in a dust-gas layer and formation of
  planetesimals in the solar nebula}.
\newblock \textit{Progress of Theoretical Physics} 69:1116--1130

\bibitem[{{Sekiya}(1998)}]{sek98}
{Sekiya} M. 1998.
\newblock {Quasi-Equilibrium Density Distributions of Small Dust Aggregations
  in the Solar Nebula}.
\newblock \textit{Icarus} 133:298--309

\bibitem[{{Sreenivasan} \& {Stolovitzky}(1995)}]{sreeni}
{Sreenivasan} KR, {Stolovitzky} G. 1995.
\newblock {Turbulent cascades}.
\newblock \textit{Journal of Statistical Physics} 78:311--333

\bibitem[{{Stepinski} \& {Valageas}(1996)}]{sv96}
{Stepinski} TF, {Valageas} P. 1996.
\newblock {Global evolution of solid matter in turbulent protoplanetary disks.
  I. Aerodynamics of solid particles.}
\newblock \textit{Astronomy and Astrophysics} 309:301--312

\bibitem[{{Stevenson} \& {Lunine}(1988)}]{stev88}
{Stevenson} DJ, {Lunine} JI. 1988.
\newblock {Rapid formation of Jupiter by diffuse redistribution of water vapor
  in the solar nebula}.
\newblock \textit{Icarus} 75:146--155

\bibitem[{{Takeuchi} \& {Lin}(2002)}]{tak02}
{Takeuchi} T, {Lin} DNC. 2002.
\newblock {Radial Flow of Dust Particles in Accretion Disks}.
\newblock \textit{Astrophysical Journal} 581:1344--1355

\bibitem[{{Tanga} et~al.(2004){Tanga}, {Weidenschilling}, {Michel} \&
  {Richardson}}]{tanga}
{Tanga} P, {Weidenschilling} SJ, {Michel} P, {Richardson} DC. 2004.
\newblock {Gravitational instability and clustering in a disk of
  planetesimals}.
\newblock \textit{Astronomy and Astrophysics} 427:1105--1115

\bibitem[{{Testi} et~al.(2003){Testi}, {Natta}, {Shepherd} \& {Wilner}}]{testi}
{Testi} L, {Natta} A, {Shepherd} DS, {Wilner} DJ. 2003.
\newblock {Large grains in the disk of CQ Tau}.
\newblock \textit{Astronomy and Astrophysics} 403:323--328

\bibitem[{{Throop} \& {Bally}(2005)}]{tb05}
{Throop} HB, {Bally} J. 2005.
\newblock {Can Photoevaporation Trigger Planetesimal Formation?}
\newblock \textit{Astrophysical Journal Letters} 623:L149--L152

\bibitem[{{Toomre}(1964)}]{toomre64}
{Toomre} A. 1964.
\newblock {On the gravitational stability of a disk of stars}.
\newblock \textit{Astrophysical Journal} 139:1217--1238

\bibitem[{{Toomre}(1981)}]{t81}
{Toomre} A. 1981.
\newblock {What amplifies the spirals}.
\newblock In \textit{Structure and Evolution of Normal Galaxies}, eds.
  SM~{Fall}, D~{Lynden-Bell}

\bibitem[{{Toschi} \& {Bodenschatz}(2009)}]{toschi}
{Toschi} F, {Bodenschatz} E. 2009.
\newblock {Lagrangian Properties of Particles in Turbulence}.
\newblock \textit{Annual Review of Fluid Mechanics} 41:375--404

\bibitem[{{Voelk} et~al.(1980){Voelk}, {Jones}, {Morfill} \& {Roeser}}]{volk}
{Voelk} HJ, {Jones} FC, {Morfill} GE, {Roeser} S. 1980.
\newblock {Collisions between grains in a turbulent gas}.
\newblock \textit{Astronomy and Astrophysics} 85:316--325

\bibitem[{{Vorobyov} \& {Basu}(2008)}]{vb4}
{Vorobyov} EI, {Basu} S. 2008.
\newblock {Mass Accretion Rates in Self-Regulated Disks of T Tauri Stars}.
\newblock \textit{Astrophysical Journal Letters} 676:L139--L142

\bibitem[{{Ward}(1976)}]{ward76}
{Ward} WR. 1976.
\newblock {The formation of the solar system.}
\newblock In \textit{Frontiers of Astrophysics}, ed. EH~{Avrett}. Harvard
  University Press

\bibitem[{{Ward}(2000)}]{ward00}
{Ward} WR. 2000.
\newblock {On Planetesimal Formation: The Role of Collective Particle
  Behavior}.
\newblock In \textit{Origin of the earth and moon, edited by R.M.~Canup and
  K.~Righter and 69 collaborating authors.~Tucson: University of Arizona
  Press., p.75-84}, eds. RM~{Canup}, K~{Righter}, {et al.} University of
  Arizona Press

\bibitem[{{Watson} et~al.(2007){Watson}, {Stapelfeldt}, {Wood} \&
  {M{\'e}nard}}]{watson}
{Watson} AM, {Stapelfeldt} KR, {Wood} K, {M{\'e}nard} F. 2007.
\newblock {Multiwavelength Imaging of Young Stellar Object Disks: Toward an
  Understanding of Disk Structure and Dust Evolution}.
\newblock In \textit{Protostars and Planets V}, eds. B~{Reipurth}, D~{Jewitt},
  K~{Keil}

\bibitem[{{Weidenschilling}(1977{\natexlab{a}})}]{weidenschilling77b}
{Weidenschilling} SJ. 1977{\natexlab{a}}.
\newblock {Aerodynamics of solid bodies in the solar nebula}.
\newblock \textit{Monthly Notices of the Royal Astronomical Society} 180:57--70

\bibitem[{{Weidenschilling}(1977{\natexlab{b}})}]{weidenschilling77a}
{Weidenschilling} SJ. 1977{\natexlab{b}}.
\newblock {The distribution of mass in the planetary system and solar nebula}.
\newblock \textit{Astrophysics and Space Science} 51:153--158

\bibitem[{{Weidenschilling}(1980)}]{weiden80}
{Weidenschilling} SJ. 1980.
\newblock {Dust to planetesimals - Settling and coagulation in the solar
  nebula}.
\newblock \textit{Icarus} 44:172--189

\bibitem[{{Weidenschilling}(1984)}]{weid84}
{Weidenschilling} SJ. 1984.
\newblock {Evolution of grains in a turbulent solar nebula}.
\newblock \textit{Icarus} 60:553--567

\bibitem[{{Whipple}(1972)}]{whi72}
{Whipple} FL. 1972.
\newblock {On certain aerodynamic processes for asteroids and comets}.
\newblock In \textit{From Plasma to Planet}, ed. A~{Elvius}

\bibitem[{{Wyatt}(2008)}]{wyatt}
{Wyatt} MC. 2008.
\newblock {Evolution of Debris Disks}.
\newblock \textit{Annual Review of Astronomy and Astrophysics} 46:339--383

\bibitem[{{Youdin}(2008)}]{houches}
{Youdin} A. 2008.
\newblock {From Grains to Planetesimals: Les Houches Lecture}.
\newblock \textit{ArXiv e-prints}

\bibitem[{{Youdin} \& {Johansen}(2007)}]{yj07}
{Youdin} A, {Johansen} A. 2007.
\newblock {Protoplanetary Disk Turbulence Driven by the Streaming Instability:
  Linear Evolution and Numerical Methods}.
\newblock \textit{Astrophysical Journal} 662:613--626

\bibitem[{{Youdin}(2004)}]{youdin04}
{Youdin} AN. 2004.
\newblock {Obstacles to the Collisional Growth of Planetesimals}.
\newblock In \textit{Star Formation in the Interstellar Medium: In Honor of
  David Hollenbach}, eds. D~{Johnstone}, FC~{Adams}, DNC {Lin}, DA~{Neufeeld},
  EC~{Ostriker}, vol. 323 of \textit{Astronomical Society of the Pacific
  Conference Series}

\bibitem[{{Youdin}(2005)}]{youd05a}
{Youdin} AN. 2005.
\newblock {Planetesimal Formation without Thresholds. I: Dissipative
  Gravitational Instabilities and Particle Stirring by Turbulence}.
\newblock \textit{ArXiv Astrophysics e-prints}

\bibitem[{{Youdin} \& {Chiang}(2004)}]{yc04}
{Youdin} AN, {Chiang} EI. 2004.
\newblock {Particle Pileups and Planetesimal Formation}.
\newblock \textit{Astrophysical Journal} 601:1109--1119

\bibitem[{{Youdin} \& {Goodman}(2005)}]{yg05}
{Youdin} AN, {Goodman} J. 2005.
\newblock {Streaming Instabilities in Protoplanetary Disks}.
\newblock \textit{Astrophysical Journal} 620:459--469

\bibitem[{{Youdin} \& {Lithwick}(2007)}]{yl}
{Youdin} AN, {Lithwick} Y. 2007.
\newblock {Particle stirring in turbulent gas disks: Including orbital
  oscillations}.
\newblock \textit{Icarus} 192:588--604

\bibitem[{{Youdin} \& {Shu}(2002)}]{ys}
{Youdin} AN, {Shu} FH. 2002.
\newblock {Planetesimal Formation by Gravitational Instability}.
\newblock \textit{Astrophysical Journal} 580:494--505

\end{thebibliography}













\end{document}